\documentclass[authoryear,3p,times]{elsarticle}

\usepackage{amssymb}
\usepackage{lineno}
\usepackage{graphicx}
\usepackage[T1]{fontenc}
\usepackage[utf8]{inputenc}
\usepackage{geometry}
\usepackage{amsmath}
\usepackage{booktabs}
\usepackage[ruled, linesnumberedhidden, shortend]{algorithm2e}
\SetAlCapHSkip{0em}

\usepackage{lscape}

\def\Cnodeput(#1)#2#3#4{\cnode(#1){#2}{#3}\rput(#3){#4}}
\usepackage[nice]{nicefrac}
\usepackage{enumitem}
\usepackage{longtable}

\newcommand{\ebbs}{\text{EBBS}}
\newcommand{\psc}{\text{PSC}}
\newcommand{\cpmc}{\text{CPMC}}
\newcommand{\ran}{\text{R1}}
\newcommand{\rann}{\text{R2}}
\newcommand{\pow}{{\cal P}(R)}

\begin{document}

\begin{frontmatter}

\title{An exact and two heuristic strategies for truthful bidding in \\combinatorial transport auctions}

\author{Tobias Buer\fnref{authorlabel}}
\address{Computational Logistics Junior Research Group, University of Bremen, Germany}
\fntext[authorlabel]{Computational Logistics -- Cooperative Junior Research Group of University of Bremen and ISL - Institute of Shipping Economics and Logistics, Wilhelm-Herbst-Str. 5, 28359 Bremen, Germany}
\ead{tobias.buer@uni-bremen.de}
\ead[url]{http://www.cl.uni-bremen.de}

\date{Received: date / Accepted: date}
\begin{abstract}
To support a freight carrier in a combinatorial transport auction, we proposes an exact and two heuristic strategies for bidding on subsets of requests.
The exact bidding strategy is based on the concept of \emph{elementary} request combinations.
We show that it is sufficient and necessary for a carrier to bid on each elementary request combination in order to guarantee the same result as bidding on each element of the powerset of the set of tendered requests.
Both heuristic bidding strategies identify \emph{promising} request combinations.
For this, pairwise synergies based on saving values as well as the capacitated p-median problem are used.
The bidding strategies are evaluated by a computational study that simulates an auction.
It is based on 174 benchmark instances and therefore easily extendable by other researchers.
On average, the two heuristic strategies achieve 91 percent and 81 percent of the available sales potential while generating 36 and only 4 percent of the bundle bids of the exact strategy.
Therefore, the proposed bidding strategies help a carrier to increase her chance to win and at the same time reduce the computational burden to participate in a combinatorial transport auction.
\end{abstract}

\begin{keyword}
combinatorial auction \sep contract procurement \sep bid generation problem \sep bundle bidding
\end{keyword}

\end{frontmatter}


\SetKwFunction{cps}{clusterPairwiseSynergies}
\SetKwFunction{c}{cluster}
\SetKwFunction{g}{generateAllElementaryRequestSets}
\SetKwFunction{e}{evaluateElementaryRequestSet}

\section{Introduction} \label{sec:intor}

\subsection{Relevance of combinatorial transport auctions} \label{sec:introduction}
In order to buy and sell services for multiple heterogenous transport requests combinatorial transport auctions are used.
On the one hand, combinatorial auctions are used by large shippers to buy their required transport services from many carriers \citep{Caplice_2003, Caplice_2006, Sheffi_2004}.
The shipper acts as the auctioneer and the carriers are the bidders in the auction.
In such a scenario, usually long-term commitments in the sense of a framework agreement are negotiated.
Real-world transport procurement auctions are voluminous, i.\,e., in a single auction some hundred requests with an annual value of several million US-Dollar are allocated.
Examples of companies that use combinatorial transport auctions for this purpose are Home Depot, Wal Mart, Procter \& Gamble, and Ford Motors \citep{Caplice_2006}.
On the other hand, combinatorial auctions can also be used by carriers to implement collaborative planning for a coalition of carriers \citep{Kopfer_1999, Krajewska_2006, Berger_2010}.
Here, the planning horizion is rather short-term.
Transport requests are exchanged between members of the coalition.
The exchange can be organized by combinatorial transport auctions and the goal of the exchange is to find an efficient allocation of transport requests to carriers.

In contrast to conventional auctions, combinatorial auctions allow bidders to submit bundle bids.
A \emph{bundle bid} (also referred to as package bid) is a bid on any subset of the tendered transport requests.
The bid price is the amount of money the bidding carrier charges to fulfill all transport requests of the subset.
Furthermore, bundle bids are all-or-nothing bids, i.\,e., a carrier either wins all requests composed in a bundle bid or none.
For this reason, carriers are safe to express their valuations for combinations of requests which usually differ from the sum of the valuations of the respective individual requests.
This is due to economies of scope and the effects of the compatibility between transport requests.
That is, the cost to fulfill a request significantly depend on the amount and characteristics of additional requests a carrier has to fulfill as well, e.\,g. the requests can be executed at lower costs, if they can be combined in a well balanced tour.
Carriers benefit from more efficient transport plans and an increased planning reliability which is enabled by bundle bids.
What is more, the shipper has a significant potential to lower his total procurement costs due to bundle bidding.

Certainly, the advantage to fully express valuations for request combinations poses a challenge for the bidders.
A bidder that does not submit bundle bids is no longer competitive.
Determining the bidder's preferences becomes more complicated, because the bidder has to evaluate a huge number of request combinations which grows exponentially with the number of tendered transport requests.
In line with this, \citet{Plummer_2003} and \citet{Elmaghraby_2004} report that real world carriers are often far from unlocking the full potential of bundle bidding.
Furthermore, the results from classical auction theory with single-item based auctions are difficult to apply to combinatorial auctions with multiple items (i.e., transport requests).
Therefore, we are interested in providing decision support for bundle bidding in a combinatorial transport auction.
We want to know:
Which bidding strategies reduce the effort to calculate preferences for all possible request combinations without lowering the chance to win in the auction?
Furthermore, how can we evaluate those bidding strategies by means of computational experiments and make them easily comparable for other researchers?

If bidding carriers would know more about bidding strategies, they could gain several benefits.
They could increase the chance of their bids to be accepted during winner determination.
At the same time, the effort to generated the bids could be reduced.
This could lower the entrance barriers for smaller carriers to participate in transport auctions.
Likewise, if the transport auction is used for collaborative transport planning within a groupage system, more and more diversified carriers are able to cooperate and it may become easier to lift synergies.
On the other hand, the shipper benefits from more bids and in particular more competitive bids.
Of course, from an economic point of view, more competitive bids finally lead to a more efficient usage of the available transport resources.

To get more insights into bidding in combinatorial transport auctions, we formulate a bid generation problem for less-than-truckload (LTL) requests and we study the interdependencies between valuations of request combinations.
Based on this, we propose three different bidding strategies.
These are applicable without knowledge about bids of rivaling carriers, i.e., we assume truthful bidding.
The performance of the strategies is evaluated by means of a computational benchmark study.
The study is setup such that it is easily repeatable by other researchers who might want to propose their own bidding strategies and compare them with the strategies proposed in this paper.

\subsection{Review of the literature and contribution} \label{sec:literature}
Two main problems in combinatorial auctions are the bid generation problem (BGP) and the winner determination problem (WDP).
In the BGP, a carrier (acting as a bidder) has to decide on which sets of requests to bid.
This includes a \emph{valuation} of possible bids and a \emph{selection} of bids that are actually submitted.
After bidding is complete the shipper (acting as an auctioneer) solves a WDP, i.e., the shipper chooses a set of winning bids from the set of all submitted bids.
The WDP is usually modeled based on the set partitioning problem or the set covering problem.
Designing a bidding strategy depends on the used auction rules and the formulation of the WDP.
Therefore, we start with a brief review of the literature on WDP in transport auctions.

The formulation and solution of variants of the WDP in transport auctions still receives the most attention in the relevant literature.
In this class of NP-hard combinatorial optimization problems, the task of an auctioneer is to decide, which of the submitted bundle bids should be accepted as winning bids such that the required transport services are procured with minimal total cost.
Integrating extensive real-world business constraints \citep[e.\,g.,][]{Sheffi_2004, Caplice_2006}, considering transport quality by means of Pareto optimization \citep[e.\,g.,][]{Buer_2010a, Buer_2010b, Buer_2014} or taking into account stochastic effects \citep{Remli_2013} are discussed.
For a literature review on winner determination problems see e.g.\ \citet{Abrache_2007}.
Typically, WDP models assume that the set of bundle bids submitted by all bidders is a given input parameter.

Compared to the number of papers on the WDP only a few papers deal with the BGP.
The following approaches deal with innovative bidding strategies on individual, non-bundled requests.
\citet{Figliozzi_2007} study a dynamic vehicle routing scenario with bidding in a competitive environment.
The sequential auction mechanism of \citet{Duin_2007} also deals with short-term excess transport demand by means of an online planning approach.
Requests are periodically announced by the shipper and carriers may bid on single requests.
\citet{Garrido_2007} propose a double auction format where shippers and carriers act as bidders at the same time.
More double auction mechanisms for transport requests are proposed by \citet{Huang_2013}.
Essentially, a multi-unit auction is performed for each request.
As the mechanisms are incentive compatible, the best strategy for each bidder is to bid truthfully.
\citet{Xu_2013} advance their double auction mechanisms in order to cope with a dynamic environment.
However, all these approaches are concerned with bidding on individual requests and do not support bundle bidding.

Bundle bidding strategies may be distinguished between strategies that are \emph{independent of the rivaling bidders' valuations} and strategies that \emph{do require information about the rivaling bidders' valuations}.
For the latter, the amount of required information may vary significantly.
While some strategies require \emph{complete information}, i.e., the preference function to calculate all bids, other estimate the prices of individual requests or a few request combinations only.

\emph{Complete information} approaches require that carriers are willing to share sensitive data (e.g.\ utilization of vehicle fleet, cost rates, or operational constraints) with the auctioneer. Alternatively, the auctioneer may estimate such information.
In the end, the bid price is calculated by the auctioneer and not by the carrier \citep{Chen_2009, Beil_2007}.
Possible scenarios are the ex ante existence of framework agreements that already specify which resources a carrier has to provide; here, the auction is used for operational fine-tuning.
Also, a setting were carriers have very high incentives to collaborate applies.
For example, because the carriers' do not represent independent companies but profit centers of a single company which are more open for sharing sensitive data, e.g.\ as considered in \citet{Krajewska_2008}.

The following bidding strategies require \emph{some information} (deterministic or stochastic) on the valuations of rivaling carriers.
\citet{Lee_2007a} propose a quadratically constrained formulation of a BGP.
The model assumes a multiple round combinatorial auction.
For each request, an ask prices announced by the auctioneer \citep[see ][]{Kwon_2005} is given.
Although the assumptions about bundle valuations are not made by the bidders, the bidders' require the ask prices from auctioneer in order to decide about their own bid.
The efficiency and effectiveness of solution techniques that compute a bundle bid are evaluated.
However, the study does not go beyond that point and does not focus on the performance of the computed bids in an auction setting.
\citet{Bichler_2009} analyse iterative combinatorial auctions with respect to the allocative efficiency and utility distribution between the auctioneer and the bidders.
Nevertheless, six different bidding strategies are presented and compared.
These are based on a generic value model and the ask prices announced by the auctioneer.
Furthermore, the used WDP assumes, at most one bid per bidder may win.
For these reasons, the focus is rather on the selection of bids as well as on the evaluation of the auction mechanism.

\citet{Triki_2014} present a stochastic optimization model that requires assumptions about the cheapest price for each subset of requests offered by some rivaling bidders.
These prices are considered stochastic.
Stochastic optimization is used to calculate a single bundle bid together with a bid price.
This bid is expected to maximize the bidder's profit.
The model may be solved repeatedly, if multiple bundle bids are required.
\citet{An_2005} propose two domain independent bundle bidding strategies.
The bidder's valuation for each item as well as her synergy value for each pair of items are given.
With it, the valuations for the most promising set of items are calculated.
While one strategy requires information on the single-item valuation of rivaling bidders, the other strategy requires no information on the valuations of rivaling bidders.
Both strategies are compared, among others, with a strategy that enumerates the complete set of bundle bids.
At this, auctions with up to 10 items are considered.

Finally, there are bidding strategies that work \emph{independent of the valuations} of rivaling carriers.
The strategies are \emph{truthful}, because the carrier bids according to her true valuations.
However, this is not always enforced by an incentive compatible auction mechanism for which truthful bidding is the dominant strategy.
Rather, it is due to the complexity of the BGP in transport auctions which also makes systematic cheating a nontrivial task.
\citet{Schoenberger_2004} study an auction-based request reallocation mechanism for a coalition of carriers.
Tendered are those requests which are not profitable for an individual carrier.
A bid is generated by solving a pickup and delivery selection problem which maximizes the carrier's profit.
All new requests in the solution are combined into a bundle bid and the charged price equals the marginal increase of the carrier's profit.
The evaluation focuses on the increase in the total profit of a coalition of carriers but not on the performance of the bidding strategy.
\citet{Song_2005} consider the procurement of full truckload pick up and delivery requests.
To generate bids, a truckload vehicle routing problem which minimizes the empty movement costs is solved.
Each tour of this solution is interpreted as a candidate bid; the bid price corresponds to the empty movement costs.
To select bids from the set of candidate bids that are actually submitted to the auctioneer, a set covering problem is solved which minimizes the sum of the prices of the selected bids and ensures that on each tendered request at least on bid is submitted.
The latter constraint might be unnecessary in other transport auction scenarios, as it is often not required the each carrier provides a bids on each request.
To evaluate the strategies, two agents compete against each other in a computational simulation of an auction with four to ten requests.
\citet{Wang_2005} explain first-order and second-order synergies between set of requests and they present the notion of a carrier's set of optimal bundle bids.
However, the synergy notions are neither implemented in the two bidding strategies and therefore not evaluated.
\citet{Chang_2009} considers the BGP  in a full-truckload scenario and models the problem as a variant of the minimum cost flow problem with pairwise synergies between arcs.
The evaluation focuses on the performance of the proposed method to solve the synergistic flow problem.
However, it is not studied how competitive the generated bids are in an auction setting, which is the focus of our paper.

The contribution of this paper is to propose three strategies for bidding in a combinatorial transport auction that require no information about the valuation of rivaling bidders.
The competitiveness of the bundle bids generated by the strategies are evaluated by means of a computational benchmark study.
In detail:

\begin{enumerate}
  \item An exact bundle bidding strategy is proposed.
        With this strategy, a bidder always achieves the same outcome as bidding on each element of the powerset of the set of requests.

  \item A first heuristic bidding strategy is presented that restricts the search in the bid space by using pairwise synergies between requests based on modified Clarke and Wright savings values.

  \item A second heuristic strategy uses the capacitated p-median problem as a clustering approach to identify promising request combinations and to restrict the search in the bid space even stronger.

  \item None of the strategies requires information about valuations of rivaling carriers.
        This is advantageous, because it is very difficult for a bidder to estimate what prices rivaling carriers may charge for which request combinations.

  \item The computational evaluation by means of test instances (see electronic appendix) is set up such that other researcher can easily compare their own strategies with the results of the strategies presented in this paper.
\end{enumerate}

\subsection{Organization}
The remaining paper is organized as follows.
In Section~\ref{sec:bgp}, we formulate a basic bid generation problem from the auctioneer and from the carrier point of view.
In Section \ref{sec:es} we present an exact bidding strategy for a carrier.
The heuristic bidding strategies are introduced in Section~\ref{sec:hs}.
The performance of the three bidding strategies is evaluated by means of a computational study in Section~\ref{sec:evaluation}.
Section \ref{sec:conclusion} concludes the paper and gives an outlook on future research possibilities. 
 
\section{A bundle bid generation problem} \label{sec:bgp}
A bundle bid generation problem (BGP) has to be solved by each carrier who participates in a combinatorial transport auction.
A carrier $c$ has to decide on which subsets of transport requests to bid and which price to charge.
The goal of carrier $c$ is to maximize her revenue won in the auction, i.e., the sum of prices of her bundle bids that are accepted by the auctioneer.
What makes the BGP difficult is the distributed nature of the problem.
On the one hand, there are \emph{rivaling carriers} that submit competing bundle bids whose composition and prices are unknown to carrier $c$, i.e., there is  asymmetric information.
On the other hand, the authority to decide which bids to accept as winning bids lies not with carrier $c$, but with the auctioneer, i.e., there are multiple decision makers with a distributed decision-making authority.

\subsection{Assumptions on bidding and winner determination in a combinatorial transport auction} \label{sec:assumptions_auction}
\begin{figure}
\begin{center}
\includegraphics{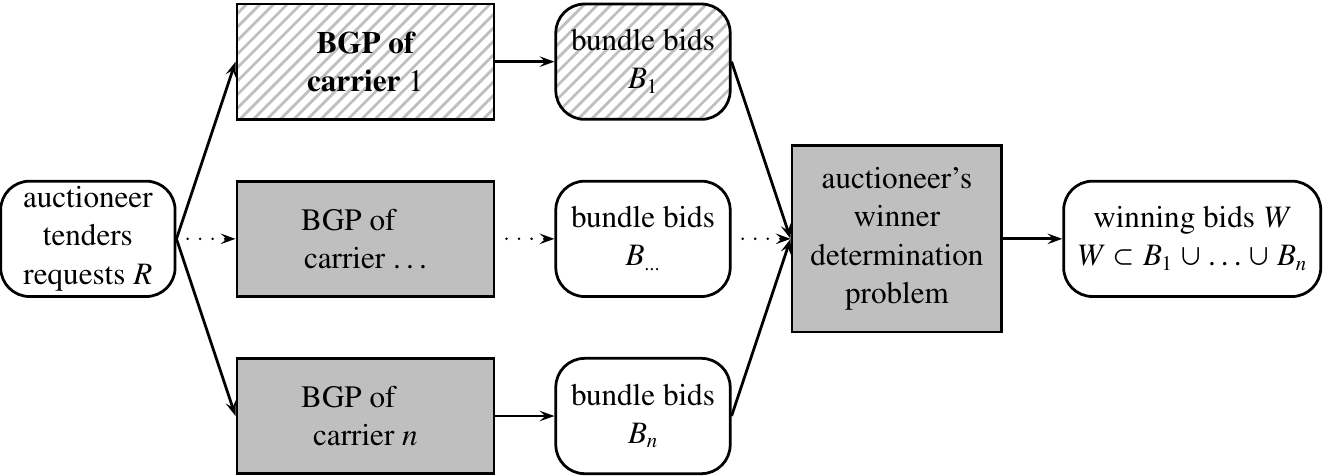}
\end{center}
\caption{Selected problems and decision makers in combinatorial transport auctions.} \label{fig:process}
\end{figure}

The process of a transport auction is shown by Figure~\ref{fig:process}.
Rectangles represent optimization problems, rounded rectangles show problem data, and edges represent input-output relations.
A directed edge from a data box to a problem box indicates that the data is required as input parameter to solve the problem.
Vice versa, an edge from a problem box to a data box indicates that the solution of the optimization problem provides the data.
So, the auctioneer announces the set $R$ of tendered requests which is used as input data by the bidders.
Each of the $n$ bidders solves her BGP, the outcome for each bidder $i$ is a set $B_i$ of bundle bids.
To decide which bundle bid is a winning bid, the auctioneer solves a WDP; the solution of the WDP is a set $W$ of winning bids, which is a subset of the set $B := B_1 \cup \ldots \cup B_n$ of all bids in the auction.

We take the point of view of a single carrier and study her bidding strategy, i.e., the way she solves the BGP.
The hatched area in Figure~\ref{fig:process} highlights the problem and the data which are controlled by this carrier.
The main problem is of course, that the success of the bidding strategy depends both on the rivaling bids and on the winner determination decision of the auctioneer, which both lie not in the circle of influence of the carrier at hand.

With respect to the auction setting, we assume the following.
\begin{itemize}
    \item Bundle bidding is allowed.
        A bundle bid $b$ is defined as a triple $(c_b, p_b, R_b)$.
        The carrier $c_b$ offers to perform the set $R_b$  of requests for the price $p_b \in \mathbb{N}$.
        The set $R_b$ is any non-empty \emph{subset} of the set of tendered requests ($R_b \subseteq R, R_b \neq \emptyset$).
        Let $\pow$ denote the powerset of $R$ without the empty set.

    \item A carrier may bid on each element of $\pow$.
         The cardinality of $\pow$ is $2^{|R|}-1$.

    \item Bidders bid truthfully.
        The price of a bundle bid reflects the true valuations of the bidder and there is \emph{no} strategic bidding.
        From this follows, that each carrier looks only on her own and she makes no assumptions on the bids of rivaling carriers.

    \item Bidders submit OR-bids.
        \citet{Nisan_2000} and \citet{Day_2009} discuss several bidding languages for combinatorial auctions.
         Here, a bidding language is characterized by logical connections between the bundle bids like OR or XOR, which allow the bidders to express different preference for combinations of bundle bids.
         We use OR-bids, i.e., the auctioneer is allowed to select any subset of the submitted bids.

    \item The carriers use cost plus pricing.
        The price of a bundle bid is proportional to the costs of performing the requests.
        Without loss of generality, we assume a markup of zero percent, i.e., the cost for performing a set of requests are equal to the price charged by a carrier to execute these requests.

    \item The WDP of the auctioneer is formulated as the set covering problem. It is given by formula (\ref{eq:wdp}).
        \begin{align}
          \min_{W \subseteq B} f^{a} = \sum_{b \in W}  p_b \quad \text{ such that } \quad \bigcup_{b \in W} R_b & = R \label{eq:wdp}
        \end{align}
        The auctioneer's task is to select a minimum cost set $W$ of winning bids $(W \subseteq B)$ that \emph{covers} each tendered request $r \in R$ at least once.
        A set $W$ of bids \emph{covers} a request $r \in R$ if it contains at least one bid $b \in W$ with $r \in R_b$.

        For the transport auction scenario at hand we prefer a WDP-formulation based on the set covering model to other formulations based on set packing or set partitioning models.
        WDP formulations based on the set packing model \citep[see, e.g.][]{Vries_2003} are better suited for scenarios where the auctioneer wants to sell goods, rather than to buy goods like in the scenario at hand.
        WDP formulations based on the set partitioning model require that each transport request is part of \emph{exactly} one winning bundle bid.
        However, given a set of bundle bids $B$, the total cost of the optimal set $W$ of winning bids under the set covering formulation is guaranteed to be \emph{lower or at most equal} than the total cost of the optimal set partitioning solution for $B$ \citep{Vries_2003, Song_2005, Buer_2010b}.
        This is clearly preferred by the auctioneer.
\end{itemize}

\subsection{Assumptions on the tendered transport requests and the bidding carrier} \label{sec:assumptions_transport}
We consider a shipper whose customers have to be serviced from the shipper's warehouse by LTL requests.
With respect to the structure of the transport requests and the carrier at hand, our transport scenario is based in the well-known capacitated vehicle routing problem (CVRP) with the following modifications and additional assumptions.
\begin{itemize}
  \item The shipper operates a single warehouse $w$ from which a set $V$ of customers has to be served.

  \item The relevant carrier operates a fleet of homogeneous vehicles.
        The capacity of each vehicle is $cap$.

  \item The shipper tenders a set $R$ of requests.
        Each request $r \in R$ requires a pickup of load at the shipper's warehouse $w$ and its delivery to a customer location $i \in V$.
        We assume there is exactly one request for each customer, i.e., $|V|=|R|$.
        Therefore, requests may be also identified in terms of the involved customer location.

  \item Like in the CVRP, we assume LTL requests, i.e., each request $r \in R$ includes a load $l_r$ ($0 \leq l_r \leq cap$) which has to be fulfilled by a single vehicle.
      However, unlike the CVRP, a vehicle has to pickup the load at the shipper's warehouse $w$ and not at the carrier's depot.

  \item Each tour $(d,w,\ldots,d)$ of a vehicle starts and ends at the home depot $d$ of the carrier.
        Furthermore, a vehicle drives immediately to the warehouse $w$ after leaving the depot to pickup the loads required to service the demands of the customers in the tour.
        In each tour, the home depot is visited twice, the warehouse is visited once, and each customer is visited at most once.

  \item The carrier estimates costs by focusing on the tendered requests only, existing requests and possible future requests by other shippers which may be acquire during the planning period are not considered.
\end{itemize}

Combinatorial transport auctions are often used to procure framework contracts for a longer period of several month.
In such a case the tendered requests may be simply interpreted, for example, as recurring requests during the planning period.

With respect to the rivaling carriers, we make the same assumptions.
Therefore, the competitive differences of the carriers are only caused by their different depot locations in relation to the shipper's warehouse and the shipper's customers locations.
For our computational study, we consider these aspects during the generation of test instances as described in Section~\ref{sec:instances}.
Due to the instance generation procedure, we can simplify the problem by setting the distance of the shipper's warehouse and the carrier's depot at hand to zero.
Without loss of generality, this allows the use of sophisticated standard solution procedures which eases the computational experiments.

\subsection{Bid generation problem of an omniscient carrier} \label{sec:bgp_omniscient}
Provided that the carrier at hand knew the bundle bids submitted by her rivaling carriers she could examine the chances of winning in the auction by means of the model given by the bid generation problem of an omniscient carrier (BGPO) as defined in (\ref{eq:min_cost}) -- (\ref{eq:cover}).
Although it is not possible for the carrier to use this model directly -- due to information asymmetry and lacking decision authority -- we use to clarify the optimization problem and to evaluate the proposed bidding strategies (cf. Section~\ref{sec:evaluation}).

For the BGPO, the set of bundle bids $B$ is partitioned into two subsets $B^c$ and $B^r$ with $B^c \cup B^r = B$ and $B^c \cap B^r = \emptyset$.
Let $B^c$ and $B^r$ denote the set of bids submitted by carrier $c$ and by all rivals of carrier $c$, respectively.
\begin{alignat}{2}
\label{eq:min_cost}
&\text{lex} \min_{W \subseteq B} f^{a} = \sum_{b \in W}  p_b,  \\
\label{eq:max_rev}
&\text{lex} \max_{B^c} f^{b} = \sum_{b \in B^c \cap W} p_b,  \\
\label{eq:cover}
& \text{s.\,t.} \quad\quad \bigcup_{b \in W} R_b = R.
\end{alignat}
In the BGPO there are two objective functions which are lexicographically ordered.
If carrier $c$ wants to simulate the outcome of the auction and design her bundle bids appropriately, she first of all has to take into account, that the shipper will select bundle bids such that his total procurement costs are minimal (\ref{eq:min_cost}). Under this precondition, the carrier may then generate a set of bundle bids $B^c$ which maximizes her total revenue won in the auction, represented by objective function $f^{\,\text{carrier } c}$. Clearly, both decisions are interdependent. Of course, in practice the set of bundle bids $B^r$ submitted by rivaling carriers is usually unknown to carrier $c$ which makes it even more challenging to formulate an adequate optimization model for the BGP taking into account asymmetric information.

With respect to the BGPO the question is which set of bids $B^c$ maximizes the winning revenue by carrier $c$?
Under the assumption of truthful bidding, the best chance to come out on top of the bids of rivaling carriers is simply to submit all possible bundle bids (\emph{brute-force strategy}), i.e., a bid on each request combination in $\pow$.
This strategy, however, is computational infeasible even for a small number of tendered request, because to calculate the bid price a NP-hard vehicle routing problem has to be solved for all $2^{|R|}-1$ request combinations.

\section{An optimal bidding strategy based on elementary request combinations} \label{sec:es}
We propose a bidding strategy denoted as \emph{elementary bundle bid search} (EBBS).
The outcome of EBBS is denoted \emph{optimal} (or exact), because it provides the same result as a brute-force strategy that bids on each request combination in $\pow$.

\subsection{Bidding strategy and bidding space} \label{sec:assumptions}
A bidding strategy $\phi$ in a combinatorial auction is a method that generates a set $B$ of bundle bids.
Considering a truthful carrier $c$, a bidding strategy $\phi^c$ is a function on the set $R$ of requests which outputs a set $B^c$ of carrier's $c$ bundle bids $b := (c, R_b, p_b), b \in B^c$.
At this, $R_b$ is a subset of $R$ and the bid price $p_b$ is calculated according to a cost function $p(R_b) \in \mathbb{N}$.
Thus, the bidding function of $c$ may be formally defined as in (\ref{eq:phi}).
For future reference, we omit the superscript $c$ for ease of notation because we focus on the individual carrier $c$ only.
\begin{align} \label{eq:phi}
  \phi^c : R \rightarrow \{b \in B^c | b = (c, R_b, p_b) \text{ with } R_b \in \pow \text{ and } p:R \rightarrow \mathbb{N}\}.
\end{align}

The bidding strategy $\phi$ generates atomic bids only \citep{Nisan_2001}.
Because the winner determination problem assumes OR-bids, additional logical dependencies between bundle bids do not have to be modeled.
That is, the auctioneer can accept any subset of the submitted bundle bids; on the other hand, a bidder cannot enforce logical constraints between atomic bids in the fashion of 'if you accept one of these bids then you cannot accept that bid.'

The bid space is $\pow \times \mathbb{N}$ with a cardinality of $|\pow|$.
There will be at most one bid on each request combination of $\pow$.
In particular, it is not reasonable for a carrier to generate multiple bundle bids with different prices on the same request combination $A, A \in \pow$.
As each of these bids offers an identical set $A$ of requests, the bid with the lowest price dominates all other bids on $A$.
Dominated bids will never be part of an optimal solution of the WDP.

The carrier uses her cost function $p(A)$ in order to calculate a bid price for $A \in \pow$.
The price charged for executing the requests $R$ is equal to the minimum cost solution of the VRP described in Section~\ref{sec:assumptions_transport}.
Form this bid price function, three plain bid strategies arise naturally. Given a set $R$ of tendered requests, a truthful bidder could:
\begin{enumerate}
  \item solve the VRP for $R$ and submit a single bid on $R$ only,
  \item solve the VRP for $R$ and submit a bid on each generated tour, or
  \item solve the VRP for each request combination in $\pow$ and submit a bid on each element of $\pow$.
\end{enumerate}

As introduced in Section~\ref{sec:bgp_omniscient}, the auctioneer minimizes his total procurement costs by solving the WDP given by formula (\ref{eq:wdp}).
The first strategy generates a single bid which is as efficient as it can be for the carrier at hand.
Nevertheless, it will be not very competitive in the WDP as it somewhat ignores the rivaling carriers which also submit bids.
In order to win any revenue in the auction, the bid of our carrier has to offer the lowest total costs for $R$ compared to all possible combinations of bundle bids which also cover $R$.
This seems highly unlikely for larger transport auctions.

The second strategy generates several bids which inhibit the same expressiveness as the single bid of the first strategy.
If all bids of the second strategy are accepted by the auctioneer, the assigned requests and the won revenue are identical.
With this in mind, the third strategy will be at least as successful as the first and the second strategy.
On the one hand, simply because the set of the generated bids are supersets of the bids generated by the fist and the second strategy.
On the other hand, because it generates more bids which offer a higher chance to be a good match with bids of rivaling carriers.
Of course, the computational effort for solving $2^{|R|}-1$ vehicle routing problems is prohibitive high considering the number of requests in many real world auctions.
In the following, we develop an approach which guarantees the same results as the brute force strategy but generates significantly less bids.

\subsection{Relations between the valuations of sets of requests}

Some terminology with respect to the relationships between costs for performing sets of requests are introduced.
First, performing an additional request always increases (or at least does not decrease) the total costs of a carrier.
Let $A$ and $B$ be two sets of requests ($A,B \subseteq R$).
If $A \subseteq B$ it follows that $p(A) \leq p(B)$.
This relationship is denoted as \emph{free disposal}. The term originates from the point of view of an auctioneer that does not have to pay a fee for disposing purchased items wherefore the auctioneer's utility of receiving additional items given the same price (ceteris paribus) never decreases. Due to the pricing function $p$, free disposal is a guaranteed characteristic of the scenario at hand.

Second, \citet{Nisan_2000} introduced subadditive, superadditive, and additive valuations between \emph{disjoint} sets of items in combinatorial auctions.
We apply their terminology with respect to the valuation of request combinations from the point of view of a single carrier.
Let $A$ and $B$ be two \emph{disjoint} subsets of requests ($A,B \subseteq R, A \cap B = \emptyset$).
\begin{itemize}
  \item If $p(A \cup B) < p(A) + p(B)$, the valuation of $A$ and $B$ is strictly cost \emph{subadditive}.
  \item If $p(A \cup B) > p(A) + p(B)$, the valuation of $A$ and $B$ is strictly cost \emph{superadditive}.
  \item If $p(A \cup B) = p(A) + p(B)$, the valuation of $A$ and $B$ is cost \emph{additive}.
\end{itemize}

\begin{figure}[htb]
\begin{center}
\includegraphics{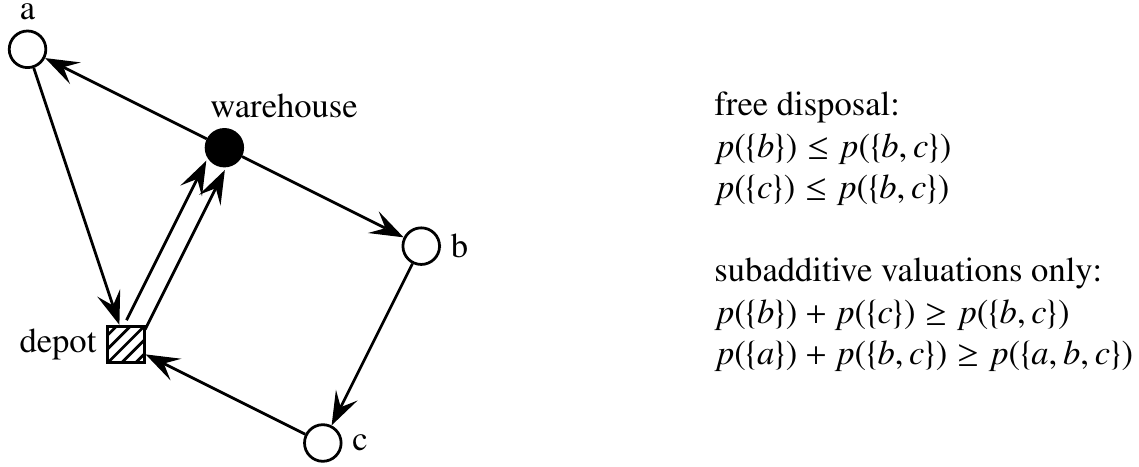}
\end{center}
\caption{Example of relations between the valuations of request combinations} \label{fig:valuations}
\end{figure}

Under the conditions of the vehicle routing problem at hand, there are only additive and subadditive valuations between sets of requests but no superadditive valuations.
Consider Figure~\ref{fig:valuations} with three requests $a,b,c$ represented by three customers nodes.
Let $A$ and $B$ be two sets of requests with $A=\{a\}$ and $B = \{b,c\}$.
It is clear, that $p(\{a, b, c\})$ is either smaller than $p(\{a\}) + p(\{b, c\})$, e.g., when all three request can be performed in a single tour and a depot edge can be removed.
Or, if two tours are required to perform the requests due to capacity constraints, the total costs are unchanged.
Therefore, $p(\{a, b, c\}) > p(\{a\}) + p(\{b, c\})$ is not possible, i.e., the valuation is not superadditive.

\subsection{Elementary request combinations and elementary bundle bids}
We show that given free disposal and the absence of superadditive valuations, it is sufficient to bid on subadditive request combinations.
We start by introducing the concept of elementary request combinations.
A subset $E$ of the set of requests $R$ is denoted as \emph{elementary} if this set of requests can be performed by at most one vehicle, i.e., $\sum_{r \in E} d_r \leq cap$ with $E \in \pow$.
The set of all elementary request combinations is:
\begin{align}
  \mathcal{E} := \{E \in \pow \, | \sum_{i \in E} d_i \leq cap \}.
\end{align}
Consequently, the set of \emph{non-elementary request combinations} is defined as
\begin{align}
  \overline{\mathcal{E}} := \pow \setminus \mathcal{E}.
\end{align}
A bundle bid $b = (p_b,c_b,R_b)$ on an elementary request combination ($R_b \in \mathcal{E}$) is an \emph{elementary bid}. A bid on a non-elementary request combination in $\overline{\mathcal{E}}$ is a \emph{non-elementary bid}.
Note, an elementary request combination for one carrier needs not to be elementary from another carrier's point of view.

All elementary request combinations exhibit subadditive valuations.
On the other hand, non-elementary request combinations exhibit additive valuations.
Superadditive valuations are not present in the scenario at hand.
Therefore, an auctioneer is able to infer the prices of all non-elementary bids of a carrier given that the carrier submits all elementary bundle bids.
To infer the price for performing a set $R'$ of requests ($R' \subseteq R$) the auctioneer has to select a cost minimal set $W$ of bids whose request combinations is a partition of $R'$, i.e., the auction has to solve the following set partitioning problem:
\begin{align}
  \min \sum_{b \in W}  p_b \quad \text{ such that } \quad \bigcup_{b \in W} R_b & = R' \quad \text{ and } \quad \bigcap_{b \in W} R_b = \emptyset
\end{align}
Table~\ref{tab:inferredprices} shows an example.
The auctioneer tenders four requests $a,b,c$, and $d$; seven subadditive bundle bids haven been submitted by a carrier, the required relative capacity (with $cap=1$) is given in the second column and the bid price is given in the column $p(R_b)$.
Now, the auctioneer can infer from these seven subadditive bids the prices of all remaining bids by solving set partitioning problems.

\begin{table}
\begin{center}
\caption{Example of inferred information on bundle bids} \label{tab:inferredprices}
\begin{tabular}{llllll} \toprule
\multicolumn{3}{c}{Elementary bundle bids} &  & \multicolumn{2}{c}{Inferred bundle bids} \\ \cmidrule(lr){1-3} \cmidrule(lr){5-6}
$R_b$ & used capacity & $p(R_b)$ &  & $R'$  & inferred prices for $R'$ \\ \midrule
\{a\} & 0.8 & $p_{a}$ &  & \{a,c\} & $p_a +p_c$ \\
\{b\} & 0.2 & $p_{b}$ &  & \{a,d\} & $p_a + p_d$ \\
\{c\} & 0.5 & $p_{c}$ &  & \{c,d\} & $p_c + p_d$ \\
\{d\} & 0.7 & $p_{d}$  &  & \{a,b,c\} & $\min\{p_A+p_b + p_cC; \, p_{ab} + p_{c}\}$ \\
\{a,b\} & 1.0 & $p_{ab}$ &  & \{a,b,d\} & $\min\{p_{ab} + p_d; \, p_a + p_b + p_d; \, p_a + p_{bd}\}$ \\
\{b,c\} & 0.7 & $p_{bc}$ &  & \{a,c,d\} & $\min\{p_{ab} + p_d; \,  p_a + p_b + p_D; \, p_A + p_{bd}\}$ \\
\{b,d\} & 0.9 & $p_{bd}$ &  & \{b,c,d\} & $\min\{p_{ab} + p_{d}; \, p_{a} + p_{b} + p_{d}; \, p_{a} + p_{bd}\}$ \\
 &  &  &  & \{a,b,c,d\} & $\min\{p_a+ p_b + p_c + p_d; \, \ldots; \, p_a + p_c + p_{bd} \}$ \\ \bottomrule
\end{tabular}
\end{center}
\end{table}

For these reasons, it is sufficient if the bidding strategy $\phi$ of a carrier takes into account all elementary request combinations $\cal E \subset \pow$ only, because the auctioneer is able to infer the bid prices of all non-elementary bids which is always implicitly done during winner determination.
That is, a bid strategy can focus on those request combinations which can be executed by a single vehicle.
This brings up another question.
Do we have to bid on each elementary request combination or can we exclude additional request combinations a priori, e.g., is it reasonable to focus on request combinations which are performed in a highly efficient tour?

\subsection{Focusing solely on efficient tours is not reasonable}
The question is whether it is necessary to compute all elementary bids in order to achieve an \emph{optimal} solution of the BGP or does it suffice to submit a subset of the elementary bids to achieve the same outcome.
The question is of practical importance due to the combinatorial nature of the BGP. For a transport auction with fifty requests auctioned and an average size five requests per elementary bid there are $\binom{50}{5}$ or more than 2,000,000 elementary bids; with an average size of ten requests per elementary bid there are already more than 10,000,000,000 elementary bids.

A natural thought to reduce the number of bids is to focus on those (elementary) request combinations that can be combined in highly efficient tours.
The idea is that an efficient tour leads to lower cost per request and therefore a bid on an efficient tour is probably more competitive, i.e., it has a higher chance to be selected as a winning bid.
On the other hand, bids on (elementary) request combinations with a low utilization of the vehicle capacity or bids on requests which are geographically distributed in an unfavorable manner are considered as inefficient.
Although these considerations are reasonable, one has to keep in mind that ultimately the auctioneer decides, by solving the WDP, which subset of the submitted bids will be the set of winning bids.
Therefore, efficiency of bids is predominantly judged by the auctioneer taking into account
\begin{itemize}
  \item the objective function and the constraints of the used winner determination problem and
  \item all bundle bids submitted by all rivaling carriers.
\end{itemize}
For the auctioneer, a bundle bid is efficient when it contributes to minimize the WDP's total procurement costs.
This, however, does not necessarily imply that low utilized pendular tour might not have better chances to be selected as a winning bid.

\begin{figure}
\begin{center}
\includegraphics{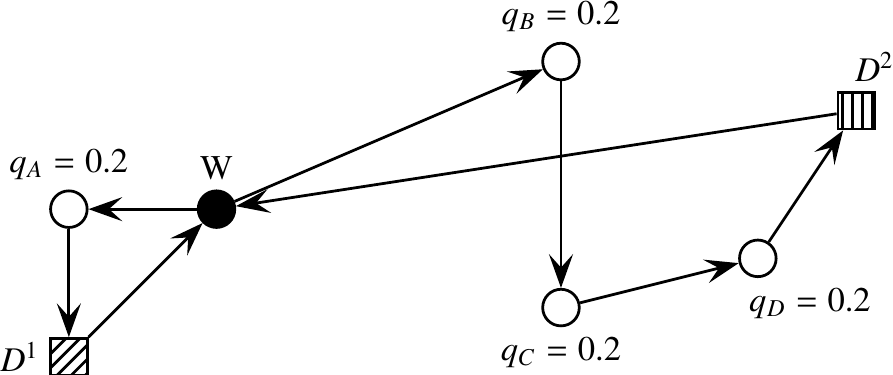}
\end{center}
\caption{Inefficient tours might still have a good chance of being selected as winning bid.} \label{fig:inefficient}
\end{figure}

Consider Figure~\ref{fig:inefficient} with two carriers $c^1, c^2$ and five bids $b^1, \ldots, b^5$. Carrier $c^2$ bids on the set of requests $\{A, B, C, D\}$ and $\{B, C, D\}$. On the other hand, carrier $c^1$ bids on $\{A\}$ and $\{A, B, C, D\}$.
Bid $b^3 = (p_3, c^1, \{A\})$ can be considered as inefficient. However, it complements very well bid $b^2$ of carrier 1. Therefore, it has a good chance to being selected as winning bid.
Assume, carrier $c^2$ with depot $D^2$ submits a bid on request $A$ which leads to a pendular tour

$b^1 = (p_1, c^2, \{A, B, C, D\})$

$b^2 = (p_2, c^2, \{B, C, D\})$

$b^3 = (p_3, c^1, \{A\})$

$b^4 = (p_4, c^1, \{A, B, C\})$

$b^5 = (p_5, c^1, \{A, B, C, D\})$

Assume, carrier $c^1$ with depot $D^1$ submits a bid on request $A$ which leads to a pendular tour with a low utilized vehicle capacity of only twenty percent. Such a pendular tour is usually considered inefficient.
From the point of view of carrier, the construction of efficient bids does not necessarily increase her chance to win in the auction.
As the carriers is not aware of the actions of her rivals, the construction of efficient bids
For a carrier, it is not a good strategy to focus on the construction of feasible bids only.

\begin{table}
  \centering

    \begin{tabular}{ccccccc} \toprule
    request & $A$ & $B$ & $A \cup B$ & $Z_1$ & $Z_2$ & $Z_3$\\ \midrule
    1 & x &  & x & x & x & \\
    2 & x &  & x & x & x & \\
    3 &  & x & x &  & & x \\
    4 &  &  &  & x & & \\ \midrule
    \multicolumn{1}{r}{$p(.)$} & 2 & 1 & 3 & 3 & 2.1 & 1.1 \\ \bottomrule
    \end{tabular}

  \caption{$A \cup B$ is redundant and less competitive than $A$ and $B$.}\label{tab:redundant}
\end{table}

From these considerations, it follows that focusing on the subset of elementary bundle bids on efficient (elementary) tours from the carrier point of view cannot be the optimal bidding strategy.
Of course, this does not exclude an adoption of these ideas within heuristic bidding strategies.
Finally, like in most auction scenarios and in accordance to many real world auctions, we assume the bidding carrier does not have knowledge about the bids of the rivaling carriers.
Therefore we cannot use this information in order to identify a subset of the set of elementary bids which leads to an optimal solution for the carrier.
For the optimal bidding strategy under the assumptions of in the transport auction at hand a carrier is required to generate all elementary bundle bids.

\subsection{A tree search approach for generating elementary bundle bids (EBBS)} \label{sec:ebbs}
We propose a bidding strategy which is denoted as elementary bundle bidding strategy (EBBS).
EBBS consists of two phases.
An overview is given by Algorithm~\ref{alg:overview}.
In the first phase (cf. Section~\ref{sec:gen_elementary}), a set of all elementary request combinations is generated.
In the second phase (cf. Section~\ref{sec:bidprice}), the price of each request combination is calculated.

\begin{algorithm} \label{alg:overview}
\caption{Elementary bundle bidding strategy (EBBS) -- Overview}

\SetAlgoLined
\DontPrintSemicolon
\SetKwFunction{c}{cluster}
\SetKwFunction{g}{generateAllElementaryRequestCombinations}
\SetKwFunction{e}{calculateBidPriceForAnElementaryRequestCombination}

\KwIn{depot node $d$ of carrier $c$, customer nodes $V$, demand $d_i, \, \forall i \in V$
 }
\BlankLine
set of bundle bids: $B \leftarrow \{\}$ \;
${\cal R} \leftarrow \g(R, d) $ \tcp*[r]{with ${\cal R} \subset \pow$}
\ForEach{$S \in {\cal R}$}{
    $p \leftarrow \e(S, d) $\;
    $b \leftarrow (c, S, p)$ \;
    $B \leftarrow B \cup \{b\}$ \;
}
\Return $B$ \;
\end{algorithm}

\subsubsection{Generation of all elementary request combinations} \label{sec:gen_elementary}

\begin{figure}
\begin{center}
\includegraphics{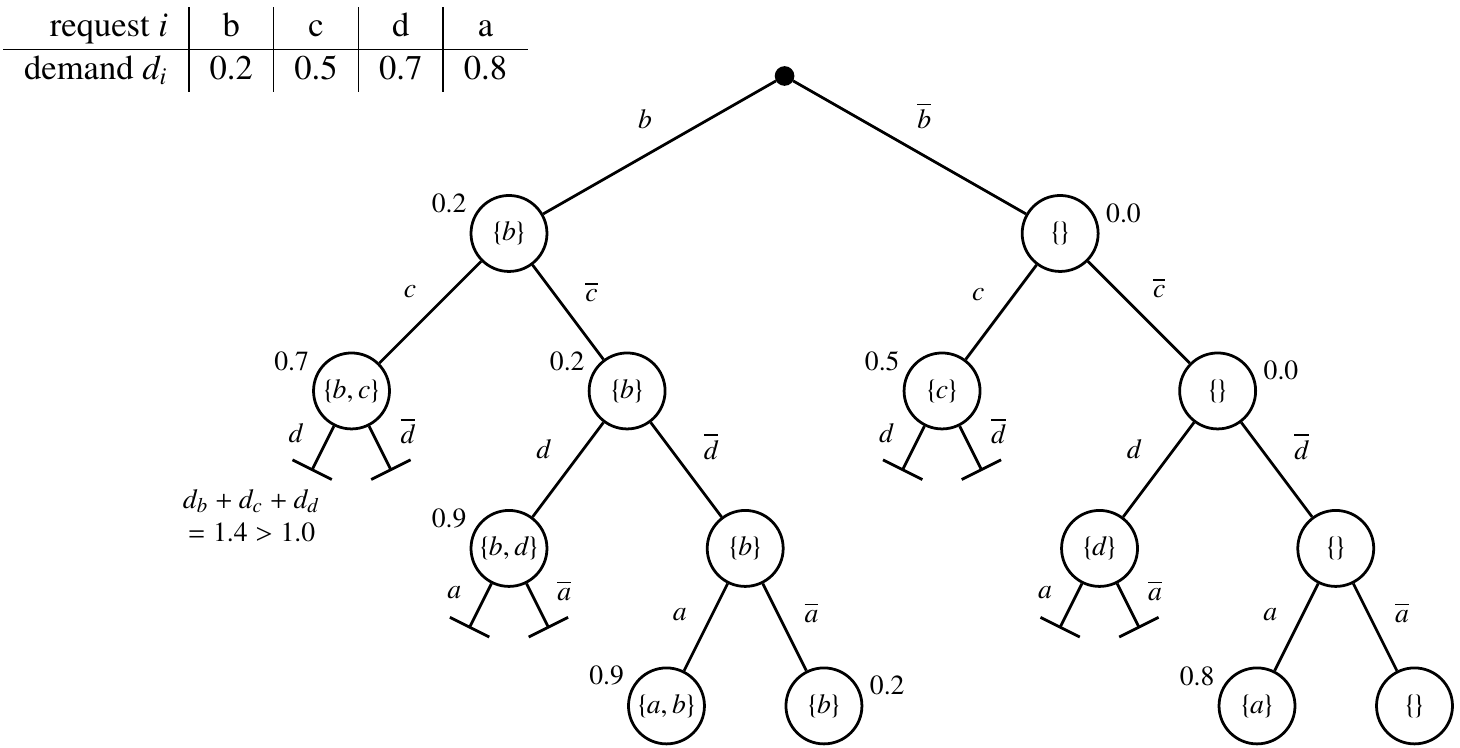}
\end{center}
\caption{Binary tree search approach to generate all elementary request combinations.} \label{fig:tree}
\end{figure}

To generate all elementary request combinations $\cal E, E \subset \pow$ binary tree search is used (cf. Figure~\ref{fig:tree}).
Assume all requests are indexed in non-ascending order of their capacity demand, i.e., the sequence of requests $(r_1,r_2,\ldots, r_{|R|})$ implies that $d_{r_i} \leq d_{r_{i+1}}$ for all $r_i \in R$. 

Each node of the binary tree represents a subset $S$ of the set of requests ($S \subseteq R$).
The root node of the binary tree represents an empty set of requests.
The number of edges $k$ on a path from the root to a given node $i$ is denoted as the level $k$ of node $i$, the root node is level 0.
The maximum level or the height of the binary tree is $|R|$.
Let $i$ be a parent node on level $k$ that represents $S$. 
The left child of node $i$ represents the set of requests $S \cup \{r_{k+1}\}$ where $r_{k+1}$ is the request indexed with $k+1$.
The right child of node $i$ represents $S$ in turn.
In other words, on level $k$ of the tree it is decided whether request $r_{k+1}$ is added to the set of requests represented by the parent node or not.

Because the requests are indexed in non-ascending order of their capacity demand, we can prune as soon as the capacity demand of the set $S$ of requests represented by a \emph{left node} exceeds the vehicle capacity, i.e., we prune if $\sum_{r_i \in S} q_i \geq cap$.
If the required capacity of the request represented by a node $l$ exceeds the vehicle capacity $cap$ the node $l$ and all its descendant nodes do not have to be visited, because they do not represent elementary request combinations.
The leaf nodes of the generated tree represent all elementary request combinations.

\subsubsection{Calculation of bid prices} \label{sec:bidprice}
According to the assumptions stated in Section~\ref{sec:assumptions} (in particular, depot $d$ equals warehouse $w$), the price of bundle bid equals the cost of performing all the requests of the bid.
By definition, elementary request combinations can be performed in a single tour, therefore we are not required to solve a vehicle routing problem but a computationally less challenging traveling salesman problem.
The problem of calculating a bid price $p$ for an elementary set of requests $S \in \cal E$ is to find the traveling salesman tour for $S$ of minimum cost.

\section{Heuristic bidding strategies based on promising sets of elementary transport requests} \label{sec:hs}
Although intelligent transportation decision support systems disseminate, the optimal bidding strategy remains computationally challenging with an increasing number of tendered requests.
Therefore, we propose two additional heuristic bidding strategies.
They are denoted as \emph{pairwise synergies based request clustering} (\psc) and \emph{capacitated p-median based request clustering} (\cpmc).
Both seek to identify a subset of elementary bids which are \emph{promising}.
Section~\ref{sec:heuristicFramework} describes the common elements of both bidding strategies, while Section~\ref{sec:pairwise} presents \psc{} and Section~\ref{sec:pmedian} describes \cpmc{}.
Two random strategies are introduced in Section~\ref{sec:random} for evaluation purposes.

\subsection{General procedure} \label{sec:heuristicFramework}
The basic idea of the heuristic bidding strategies is to generate only those elementary request combinations that are \emph{supersets} of \emph{promising} request combinations (see Algorithm~\ref{alg:ghbs} for an overview).
The bid price calculation is identical to \ebbs{}, however, the generation of bundle bids differs for the heuristic strategies.
A set $\cal P$  of promising elementary request combinations ($\cal P \subset E$) is identified by first calling the procedure \c.
The following steps slightly differ compared to the procedure of \ebbs{} in Section~\ref{sec:ebbs}:
\begin{itemize}
  \item The procedure \g is called for each $S \in \cal P$.
  \item $S$ is used as the root node of procedure \g.
  \item The outcome of \g is a set $T$ of requests that is a superset of $S$ ($T \supset S, T \in \cal E$).
  \item The set union of all generated request combinations is denoted as $\cal R$ ($\cal R \subset E$).
\end{itemize}

\begin{algorithm}
\caption{General heuristic bidding strategy} \label{alg:ghbs}

\SetAlgoLined
\DontPrintSemicolon

\KwIn{depot $d$ of carrier $c$, warehouse $w$, set of requests $R$, vehicle capacity $cap$, demand $d_i, \, \forall i \in R$
 }
\BlankLine
set of elementary request combinations ${\cal R} \leftarrow \{\}$ \;
identify promising request combinations: \quad ${\cal P} \leftarrow \c{d,w,cap, R} $ \;
\ForEach{$S \in {\cal P}$}{
    ${\cal R'} \leftarrow \g(d,w,S)$ \;
    ${\cal R} \leftarrow {\cal R} \cup {\cal R'}$ \;
}
bundle bids $B \leftarrow \{\} $ \;
\ForEach{$S \in {\cal E}$}{
    $p \leftarrow \e(d, S) $\;
    $B \leftarrow B \cup \{b\}$ with $b \leftarrow (p,c,S) $ \;
}
\Return set $B$ of bundle bids submitted by carrier $c$ \;
\end{algorithm}

The set of bundle bids generated by Algorithm~\ref{alg:ghbs} is always a subset of the set of bundle bids generated by \ebbs.
In the following Sections~\ref{sec:pairwise}, \ref{sec:pmedian}, and \ref{sec:random}, respectively, we discuss different implementations of the procedure $\c$ which identifies a promising set ${\cal P}$ of requests.

\subsection{Pairwise synergies based request clustering (\psc)} \label{sec:pairwise}
The core of the strategy \psc{} is to identify \emph{pairs} of requests that offer high synergies compared to other pairs of requests.
The idea is: if two requests $i$ and $j$ ($i,j \in R$) do not provide high synergies, then it is no-good to consider supersets of $\{i,j\}$.
Vice versa, if $i$ and $j$ already offer high synergies, then all supersets of $\{i,j\}$ are worth a closer look.
Strategy \psc{} follows Algorithm~\ref{alg:ghbs}, the \c-procedure is implemented like Algorithm~\ref{alg:hbs}.

\begin{algorithm} \label{alg:hbs}
\caption{clusterPairwiseSynergies (\psc)}
\label{alg:construct}

\SetAlgoLined
\DontPrintSemicolon
\SetKwFunction{cps}{clusterPairwiseSynergies}

\KwIn{depot $d$ of carrier $c$, warehouse $w$, set $R$ of requests, demand $d_r, \, \forall r \in R$, threshold $\alpha$
 }
\BlankLine

pairs of requests: ${\cal R}_{2} \leftarrow \{S \in 2^R : |S| = 2 \wedge \sum_{r \in S} d_r \leq cap \} $\;
\ForEach{$S \in {\cal R}_{2}$}{
    $\sigma(S) \leftarrow \underbrace{(\text{dist}_{d,i} + \text{dist}_{d,j} - \text{dist}_{i,j})}_{\text{distance saving}} \cdot \underbrace{(cap - d_i - d_j)}_{\text{idle vehicle capacity}}$ with $\quad i,j \in S, i \neq j$ \;
}
\Return $\{S \in {\cal R}_2 \, | \, \sigma(S) \in \alpha\text{-fractile of the synergy distribution}  \}$ \;
\end{algorithm}

Algorithm~\ref{alg:hbs} looks at all $\binom{|R|}{2}$ pairs of requests.
For each pair $S$ of requests a synergy measure $\sigma(S)$ is computed.
Requests offer synergies, if they complement each other nicely, i.e., if they are cost subadditive.
We say the synergy between two requests $i$ and $j$ ($\{i,j\} = S, i \neq j$) is the higher, the higher the saved travel distance due to the combined fulfillment of $i$ and $j$ is.
Saved travel distance is measured by use of the well-known \citet{Clarke_1964} savings heuristic.
Furthermore, we say the synergy of pair $\{i,j\}$ is the higher, the lower the required vehicle capacity ($cap - d_i + d_j$) is.
Given a lower utilized vehicle, ceteris paribus, the chance is higher to include more requests which complement the pair \{i,j\}.
Finally, Algorithm~\ref{alg:hbs} returns those pairs of requests that are among the $\alpha$-percent pairs with the highest synergy.

\subsection{Capacitated p-median based request clustering (\cpmc)} \label{sec:pmedian}
The second way to identify promising request combinations is borrowed from facility location.
Facility location problems address decisions about the location of facilities in a network and the allocation of demand points to these facilities.
Finding such an allocation corresponds to determining a set of clusters in a network which is why facility location problems are frequently used as clustering approaches.
A well-known problem of this class is the \emph{capacitated p-median problem} (PMP).
For an overview of model formulations and solution approaches see \citet{Reese_2006}.
The PMP considers a set of $n$ candidate points, each point has a demand.
The goal is to find a subset of $p$ points ($p \leq n$) which are denoted as medians. Each point has to be assigned to a median, such that the capacity of the median is not exceeded by the sum of the demands of the assigned points, and the total sum of distances between the points and their assigned medians is minimal.
The solution of the PMP is a partition of the $n$ points into $p$ clusters.
Here, we use the PMP as an approach to determine elementary request combinations by interpreting the PMP as follows:

\begin{alignat}{2}
\min  \quad \sum_{i \in V} \sum_{i \in V} & c_{ij} x_{ij}  \label{eq:cpmp-obj} \\
\text{subject to}\quad  \sum_{j \in V} x_{ij} &= 1, \quad && \text{for all} \quad i \in V \label{eq:cpmp-assign} \\
 \sum_{j \in V} x_{jj} &= p,  \label{eq:cpmp-p}  \\
 x_{jj} &\geq x_{ij}, \quad && \text{for all} \quad i,j \in V \label{eq:cpmp-median} \\
 \sum_{j \in V} d_j x_{ij} &\leq cap, \quad && \text{for all} \quad i \in V \label{eq:cpmp-cap} \\
 x_{ij} &\in \{0,1\}, \quad && \text{for all} \quad i,j \in V. \label{eq:cpmp-dec}
\end{alignat}

The decision variable $x_{ij}$ defined in formula (\ref{eq:cpmp-dec}) represents whether point  $i$ is assigned to the same cluster as $j$ ($x_{ij}=1$) or not ($x_{ij}=0$).
Point $i$ and $j$ stand for locations of customers ($i,j \in V$).
In other words, if $x_{ij}=1$ then the requests associated to customers $i$ and $j$ are part of the same elementary request combination.
The objective function (\ref{eq:cpmp-obj}) minimizes the total sum of the distances between medians and associated nodes.
The costs $c_{ij}$ of assigning node $i$ to node $j$ are represented by the Euclidean distance between customers $i$ and $j$.
Equation (\ref{eq:cpmp-assign}) ensures that each node is assigned to exactly one median.
Equation (\ref{eq:cpmp-p}) guarantees that there are $p$ medians and that each median is assigned to itself.
So, if $x_{jj} = 1$, then point $j$ is a median.
Points may only be assigned to medians (\ref{eq:cpmp-median}).
The capacity constraint (\ref{eq:cpmp-cap}) ensures that the total demand assigned to a median does not exceed the vehicle capacity $cap$.

PMP is NP-hard if the number $p$ of medians is a decision variable; if $p$ is given, PMP is solvable in polynomial time \citep{Garey_1979}.
Nevertheless, it is still difficult to solve even when $p$ is given.
We preset $p$ to the minimum number of medians which are required to solve PMP:
\begin{align}
  p = \Bigg\lceil \frac{\sum_{i = 1}^n d_i}{cap} \Bigg\rceil. \label{eq:p}
\end{align}

We can interpret a solution of the PMP as a set of elementary request combinations because all requests of a cluster may be performed by a single vehicle.
The heuristic bidding strategy that clusters requests by means of solving the PMP is denoted as \emph{PMP-based clustering bid strategy} (\cpmc).
Comparing the clustering approaches of the bidding strategies \psc{} and \cpmc, we expect the following differences with respect to the structure of the computed request combinations:

\begin{itemize}
  \item \cpmc{} generates a significantly smaller number of request clusters than \psc.
        The number of clusters generated by PMP is $p$, see equation (\ref{eq:p}).
        \psc, however, generates all clusters which lie in the $\alpha$-fractile according to the determined synergy distribution.
        For most values of $\alpha$ the latter will be significantly larger, since there are $\binom{|R|}{2}$ possible request clusters.

  \item \psc{} generates clusters which contain two requests.
        The number of requests in a cluster generated by \cpmc{} varies.
        Although a \cpmc{} cluster may contain a single request only, most \cpmc{} clusters will contain more than two requests which together use the available capacity $cap$ to large extend.
        This is induced due to the definition of $p$ and the solution of PMP.

 \item The sum of the demand of the requests in a cluster is higher with \cpmc{} compared to \psc.
        Therefore, \cpmc{} clusters offer a lower potential to generate additional supersets of elementary request clusters.

  \item \psc{} considers the carrier's depot $d$ and the shipper's warehouse while generating promising request clusters. \cpmc{} does not exploit this information.

  \item \cpmc{} clusters cover the tendered requests smoothly while \psc{} favors requests which are geographically closer the to carrier's depot and the shipper's warehouse.
        The PMP-clusters are a partition of the requests, i.e., the \cpmc{} places at least one bundle bid on each request.
\end{itemize}

As this discussion of the structural differences between the clustering approaches shows, both heuristic bidding strategies appear sufficiently different. It is hard to anticipate which of the strategies will be more successful in a transport auction setting why we refer to the computational study in Section~\ref{sec:evaluation}.

\subsection{Random request clustering for evaluation purposes (\ran{} and \rann{})} \label{sec:random}
For evaluation purposes we propose two random bidding strategies denoted as \ran{} and \rann{}.
\ran{} generates $|R|$ random bids.
A random bundle bid $(p_b, c_b, R_b)$ is a bid where $R_b \subseteq R$ is generated randomly but the charged price $p_b$ is calculated deterministic as described in Section~\ref{sec:bidprice}.
Starting with $R_b = \emptyset$ a request $r \in R$ is selected with probability $\frac{1}{|R|}$.
Add $r$ to $R_b$, if the vehicle capacity $cap$ suffices to perform $R_b \cup \{r\}$, update $R := R \setminus \{r\}$ and update $R_b := R_b \cup \{r\}$.
Continue extending $R_b$ until the vehicle capacity is exhausted, i.e., $d_r + \sum_{s \in R_b} d_s > cap$.
Compute the price $p_b$ for $R_b$ according to Section~\ref{sec:bidprice}.
All in all, generate $|R|$ random bundle bids.

The random strategy \rann{} follows Algorithm~\ref{alg:ghbs} like \psc{} and \cpmc{}, however, it simply uses \ran{} as an implementation for the \c-procedure.

\section{Computational benchmark study} \label{sec:evaluation}
We evaluate the effectiveness and the efficiency of the proposed bidding strategies by a computational benchmark study.
For this, we take the bidder's as well as the auctioneer's point of view.
The test setup and the used performance criteria are presented in Section~\ref{sec:eval_setup}, test instance generation is described in Section~\ref{sec:instances}, and the computational results are discussed in Section~\ref{sec:results}.

\subsection{Evaluation framework and test setup}
\label{sec:eval_setup}
The evaluation of heuristics usually involves solving a set of test instances and measuring the performance in the light of the objective function values of the computed solutions.
A solution of a bidding strategy is a set of bundle bids.
However, an objective function which can measure the quality of the generated bids \emph{directly} is not available.
Therefore, we choose an indirect approach that takes into account the independent decision authorities of bidders and auctioneer as well as the asymmetric distribution of information.

Consider Figure~\ref{fig:process} again, the evaluation is aligned according to this auction process.
There are $n$ carriers, we consider the point of view of carrier $c$. 
The test instances (see next section) include the set $R$ of tendered requests and the set $B_r$ of bundle bids submitted by the rivals of carrier $c$.
In Figure~\ref{fig:process}  they are represented by the white boxes, carrier $c$ corresponds to carrier $1$ and $B_r$ would be defined as $B_2 \cup B_3 \cup \ldots \cup B_n$.
We apply a bid strategy to compute the set $B_c$ of bundle bids of carrier $c$.
After that, the set $B := B_c \cup B_r$ of bundle bids is used as input for the winner determination problem (cf. Section~\ref{sec:assumptions_auction}) which is solved to optimality by the commercial solver CPLEX.
The resulting set $W$ of winning bundle bids ($W \subset B$) and the set $W_c$ of winning bundle bids submitted by carrier $c$ ($W_c := B_c \cap W$) are used to compute performance criteria.

Let $B^{\phi}_c$ be the set of bids generated by carrier $c$ using bidding strategy $\phi \in \Phi$, $\Phi := \{\ebbs, \psc, \cpmc, \ran, \rann\}$.
Let $W^{\phi}$ represent the set of all winning bids if carrier $c$ uses strategy $\phi \in \Phi$.
Then, the set of carrier's $c$ winning bids is $W^{\phi}_c := W^{\phi} \cap B^{\phi}_c$.

Four performance criteria $\kappa^\phi_1, \kappa^\phi_2, \kappa^\phi_3$, and $\kappa^\phi_4$ are introduced to evaluate a bidding strategy $\phi$.
From the \emph{auctioneer point of view}, $\kappa^\phi_1$ measures how much using $\phi$ increases the auctioneer's total procurement costs $f^a(W)$ compared to the exact strategy \ebbs.
Lower values of $\kappa_1^\phi$ are better.
$\kappa_1^\phi \geq 0.0$ is expected, because a heuristic bidding strategy generates a subset of the bids generated by \ebbs, which may only increase the auctioneer's total procurement cost.
\begin{align}
\kappa_1^\phi := \frac{f^a(W^{\phi}) - f^a(W^{\ebbs})}{f^a(W^{\ebbs})}. \label{eq:kappa1}
\end{align}
The remaining criteria $\kappa^\phi_2, \kappa^\phi_3$, and $\kappa^\phi_4$ consider the point of view of the \emph{bidding carrier} $c$.
$\kappa_2^\phi$ measures the \emph{used sales potential}.
The actual sales volume awarded to $c$ when using a heuristic strategy $\phi \in \Phi \setminus \{\ebbs\}$ is set in relation to the sales potential.
The carrier's sales potential is given by using the exact bidding strategy \ebbs{} and summing up the prices of $c$'s winning bids:
\begin{align}
\kappa_2^\phi := (1 + \frac{f^b(W^{\phi}) - f^b(W^{\ebbs})}{f^b(W^{\ebbs})})  \cdot  100. \label{eq:potential}
\end{align}
The success rate of a strategy $\phi$, that is the number $|W_c^\phi|$ of carrier $c$'s winning bids versus the number $|B_c^\phi|$ of carrier $c$'s bids, is measured by $\kappa_3^\phi$:
\begin{align}
\kappa_3^\phi := \frac{|W^{\phi}_c|}{|B^{\phi}_c|}  \cdot  100. \label{eq:success}
\end{align}
The ratio of the number of generated bids by strategy $\phi$ to the number of generated bids by strategy \ebbs is expressed by $\kappa_4^\phi$:
\begin{align}
\kappa_4^\phi := \frac{|B^{\phi}_c|}{|B^{\ebbs}_c|}  \cdot  100. \label{eq:ratio}
\end{align}
The computational efficiency of a strategy $\phi$ is measured by $\kappa^\phi_3$ and $\kappa^\phi_4$, while the effectiveness of $\phi$ is measured by $\kappa^\phi_1$ and $\kappa^\phi_2$ from the auctioneer's and the bidder's point of view, respectively.

Finally, some remarks on the implementation of the bidding strategies and the used hardware.
The bidding strategies are implemented in Java 1.7.
The Java code does not benefit from multicore computer architectures as parallelization of code segments was not implemented.
The prices of a bundle bid on an elementary set of requests are computed by the TSP solver Concorde by \citet{Applegate_2007}.
The winner determination problem is solved by the commercial solver IBM CPLEX Optimizer 12.5.
All computations are performed on a standard desktop computer with Intel Core i7-3770 CPU with 3.4 GHz and 16 GB of working memory using Windows 7 as operating system.

\subsection{Generation of benchmark instances} \label{sec:instances}
The bidding strategies are compared by means of benchmark instances.
An instance consists of a set $R$ of requests tendered by the auctioneer and a set $B_r$ of bundle bids of the rivals of carrier $c$.
Furthermore, the vehicle capacity $cap$ is given.
All in all, 240 instances have been generated in which between 15 and 40 requests are tendered and up to 5000 bids of rivaling carriers are present.

The set of requests is implicitly defined by a depot node $d$ and a set $V$ of customer nodes, each with a required load $l_i$ ($i \in V$).
The warehouse node $w$ and the depot node $d$ are set equal, as explained below.
The distance between two nodes is calculated as the Euclidean distance ($L_2$-norm) rounded to the nearest integer value.
The customer nodes and demands are taken from the popular instances\footnote{available at http://people.brunel.ac.uk/~mastjjb/jeb/orlib/vrpinfo.html, files vrpnc1.txt to vrpnc5.txt, vrpnc11.txt, and vrpnc12.txt} of the capacitated vehicle routing problem introduced by \citet{Christofides_1979}. The first node of a CVRP instance is used as the depot node $d$.
The subsequent $m = |R|$ nodes of a CVRP instance together with their resource demands are considered as the customer nodes of a BGP instance.

Each bundle bid $b \in B_r, b = (p_b, c, R_b)$ of a rivaling carrier is randomly constructed as follows.
First, the initial empty set $R_b := \emptyset$ of elementary requests is extended:
\begin{itemize}[leftmargin=14mm]
  \item[\textbf{Step 1.}] The predefined vehicle capacity $cap$ is randomly reduced to $cap', cap' \leq cap$, i.e., $cap$ is multiplied by a random, uniformly distributed number between zero and one.
  \item[\textbf{Step 2.}] Draw a request $r$ randomly. If $r \notin R$ and the capacity constraint $\sum_{r \in R_b} l_r \leq cap'$ is satisfied, add $r$ to $R_b$.
  \item[\textbf{Step 3.}] Repeat Step 2 until a drawn request violates the capacity constraint for the first time.
\end{itemize}
Because the actual vehicle capacity is temporary reduced from $cap$ to $cap'$ in Step 1, the method generates some elementary request combinations with a low degree of capacity utilization with respect to the vehicle capacity $cap$.
In doing so, we do not have to make assumptions about the rivaling carriers vehicle fleet, their bidding strategies or their already existing requests.

The price $p_b$ of a bundle bid $b$ is calculated as the minimum length of the traveling salesman tour that contains the warehouse $w$ and all customer nodes associated with requests in $R_b$.
The price $p_b$ is multiplied by a random number that is chosen randomly from the interval [0.7, 1.3].
By this means, we cope with different depot locations of different carriers without precisely modeling the rivaling carriers and their depot locations.
As the test results show, the generated instances are in many cases very challenging.

These rivaling bundle bids are considered as given.
For the test it does not matter whether they reflect the true valuations of the rivaling carriers or not.
Because of the random variations of the vehicle capacities and the bid prices of the rivaling carriers, it is highly likely that a bid generated by one of the bidding strategies at hand has a different bid price than a given bid by a rivaling carrier, even if both bids are on the same set of requests.
With this, the assumption that all carriers are equal except for different depot locations is adequately integrated into the test instances.
Furthermore, this is the reason why we may simplify the bid price calculation of carrier $c$ by setting the physical location of the depot of carrier $c$ equal to the physical location of the warehouse $w$.

\subsection{Results and discussion} \label{sec:results}
The detailed results for $\kappa_1$, $\kappa_2$, and $\kappa_3$ as well as run times per test instance as well as averages, median, and quartiles are reported in Table~\ref{tab:results} in the electronic appendix.

\subsubsection{Elementary bundle bidding strategy (\ebbs)}
For each of the 240 test instances we allowed a maximal computation time of 24 hours.
Within the time limit, the exact bidding strategy \ebbs{} solved 209 out of 240 instances.
Among these 209 instances, the bidder did not win any business at all in 35 cases, i.e., $f^b=0$.
That means, any truthful bidding strategy following the assumptions of Section~\ref{sec:assumptions} will not be able to win any of the tendered requests in these 35 instances, because the competition of the rivaling bids is too strong.
Furthermore, none of the proposed heuristic strategies will generate winning bids for these instances, because the set of bids generated by the heuristic strategies is always a subset of the bids generated by \ebbs{}.
Therefore, we compare the bidding strategies on those 174 instances.

As expected, the computing time of \ebbs{} reported in Table~\ref{tab:results} in the Appendix increases with an increasing number of possible elementary bundle bids per instance.
The number of potential elementary bundle bids increases with an increasing number of tendered requests as well as an increasing vehicle capacity $cap$.
Higher values of $cap$ allow more requests to be serviced by a single vehicle.
Therefore, the time limit of \ebbs{} was exceeded for some of the largest instances with 35 or 40 tendered requests and $cap \geq 60$.
Nevertheless, even some of the largest instances with 40 requests and $cap=80$ could be solved within a few minutes, probably because the distribution of the customer loads allowed \ebbs{} to prune early compared to the instances in which \ebbs{} exceeded the time limit.

\subsubsection{Comparing effectiveness of the bidding strategies}
The effectiveness of a bidding strategy $\phi$ is measured by the used sales potential $\kappa^\phi_2$.
Table~\ref{tab:potential} shows the aggregated results for each of the four heuristic strategies for $\kappa^\phi_2$.
Two effects become clear.
In contrast to both \psc{} and \cpmc{}, the random bidding strategy \ran{} is not able to utilize the potential sales volume to a noteworthy degree for the vast majority of the instances.
Therefore, the performance of \psc{} and \cpmc{} is more than simply bidding randomly.
Taking \rann{} into account, it also becomes visible, that generating random clusters of promising requests which are used as a nucleus to construct elementary bundle bids (cf. Algorithm~\ref{alg:ghbs}) is less effective.
A systematic approach to identify \emph{promising} requests clusters as used by \psc{} and \cpmc{} is meaningful and a crucial element for the design of successful bidding strategies.

\begin{table}[htb]
  \caption{Distribution of the used sales potential for 174 tested instances} \label{tab:potential}
  \begin{center}
    \begin{tabular}{lrrrr} \toprule
     & \textbf{\psc} & \textbf{\cpmc} & \textbf{\ran} & \textbf{\rann} \\ \midrule
    \# of instances $\kappa_2^\phi \geq 100\%$ & 87 & 72 & 3 & 15 \\
    \# of instances $\kappa_2^\phi \geq 75\%$ & 136 & 103 & 3 & 21 \\
    \# of instances $\kappa_2^\phi \geq 50\%$ & 160 & 122 & 5 & 35 \\
    \# of instances $\kappa_2^\phi < 50\%$ & 14  & 52 & 169 & 139 \\
    \# of instances $\kappa_2^\phi = 0\%$ & 6 & 24 & 148 & 76 \\ \bottomrule
    \end{tabular}
  \end{center}
\end{table}

\begin{figure}[htb]
  \centering
  \includegraphics{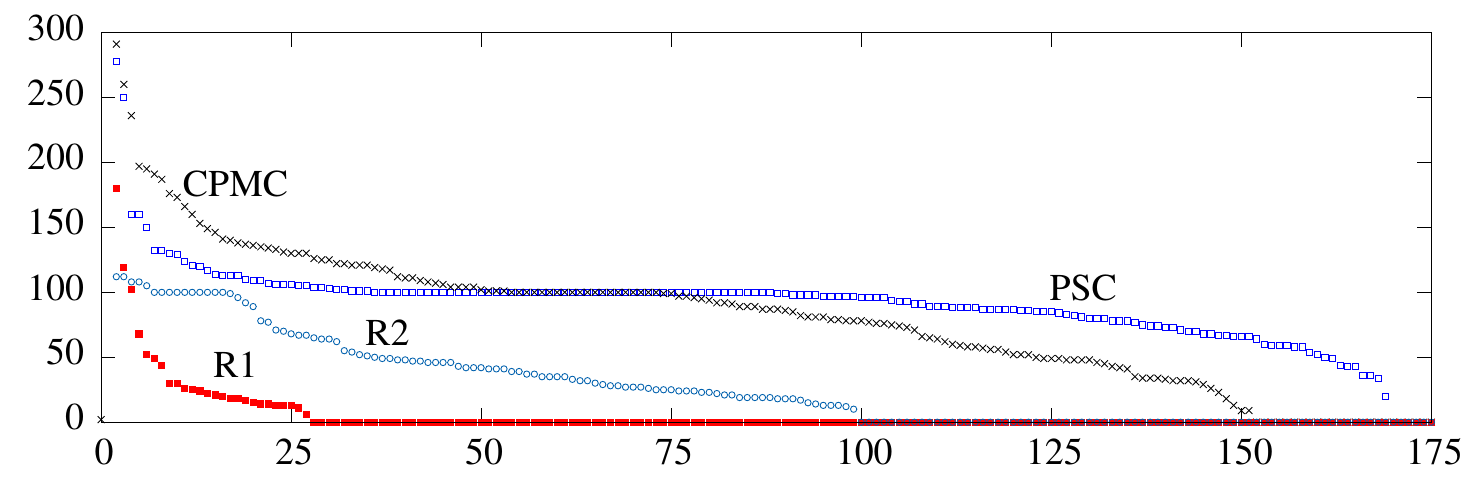}
  \caption{Comparing the used sales potential $\kappa_2^\phi$ in percent for $\phi = \psc{}, \cpmc, \ran, \rann$}\label{fig:potential}
\end{figure}

With respect to comparing the effectiveness of \psc{} and \cpmc{}, \psc{} seems to outperform \cpmc{} at least on the coarse categories given in Table~\ref{tab:potential}.
To provide more insights, we visualize the individual $\kappa^\phi_2$ values for each of the 174 instances in Figure~\ref{fig:potential}.
The chart shows, for each bidding strategy $\phi$, the obtained $\kappa^\phi_2$ values in descending order.
Both \psc{} and \cpmc{} achieve a sales potential of for more than 100\% for several instances ($\kappa_2^{\psc}, \kappa_2^{\cpmc} > 100$).
How is this possible and is it a good signal?

Recall, \ebbs{} is denoted as an exact bidding strategy because it guarantees the same outcome as bidding on each request combination in $\pow$ and not because it guarantees the maximum value for $f^b$.
A heuristic bidding strategy $\phi$ that generates a subset of all elementary bundle bids may achieve higher values for $f^b$
than the exact bidding strategy \ebbs{}.
\ebbs{} generates the most competitive bids with respect to the request combinations and the charged price.
Depending on the set of rivaling bids, however, some of the \ebbs{}'s bids could win even if higher prices are charged.
Therefore, the heuristic strategies might generate a higher sales volume compared to bidding on each request combination which implies $\kappa_2^{\phi} > 100\%$.

With this in mind, however, a bidding strategy that generates values of $\kappa_2^\phi \geq 100$ percent is double-edged.
On the one hand, we prefer strategies that generate higher sales volumes $f^b$ and therefore higher values of $\kappa_2^{\phi}$ are always better.
On the other hand, there is no free lunch.
Shifting from the static view on the issue to a dynamic view, the portion of sales which exceeds the sales potential, i.e., $\kappa_2^{\phi} - 100 > 0$, signals a competitive gap.
This gap is not earned systematically but based on fortunate circumstances with respect to the actual composition of the rivaling bundle bids which cannot be influenced by the bidder.
It also indicates the auctioneer overpays.

Looking at Figure~\ref{fig:potential} the strategy \psc{} is more robust than \cpmc{}.
\psc{} uses the sales volume more frequently, i.e., it achieves values of $\kappa_2^\phi$ near hundred percent with a higher frequency.
Both methods solve a roughly equal number of instances with a potential of hundred or more percent.
Nevertheless, the \cpmc{}-graph declines faster than the \psc{}-graph.
All in all, for about half of the tested instances, both \psc{} and \cpmc{} achieve exactly the same or an even superior outcome---with the discussed limitations---as \ebbs{}.

\subsubsection{Comparing efficiency of the bidding strategies}
The strategies \ebbs{}, \psc{}, \cpmc{} significantly vary with respect to the computational times.
Over all 209 solved instances (including the 35 instances with $f^b=0$), the \emph{median} runtime for \ebbs{}, \psc{}, \cpmc{} is 152 seconds, 9 seconds, and 3 seconds, respectively.
However, the \emph{averaged} runtime for \ebbs{}, \psc{}, \cpmc{} is 153 minutes, 20 minutes and 0.75 minutes, respectively.
Strategy \cpmc{} is an order of magnitude faster than \psc{}, and \psc{} is an order of magnitude faster than \ebbs{}.
This effect may be attributed almost exclusively to differences in the number of generated bundle bids.
The average success rates are $\kappa^{\ebbs}_3=0.035$ percent, $\kappa^{\psc}_3=0.081$ percent, and $\kappa^{\cpmc}_3=5.72$ percent.
Therefore, the heuristic strategies are a serious alternative to \ebbs{} if the trade-off between used sales volume and runtime are considered.

\begin{figure}%
\centering
\parbox{0.45\textwidth}{\includegraphics[width=7cm]{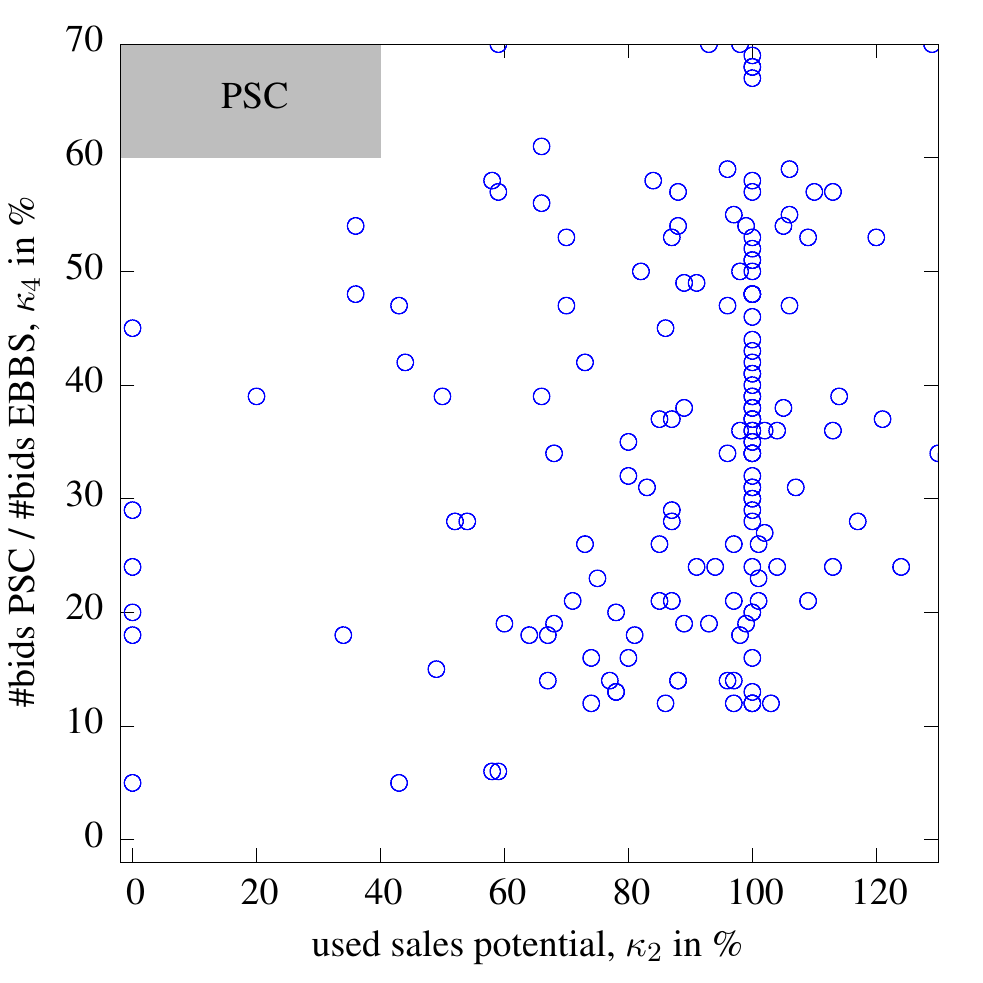}}%
\qquad
\begin{minipage}{0.45\textwidth}%
\parbox{0.45\textwidth}{\includegraphics[width=7cm]{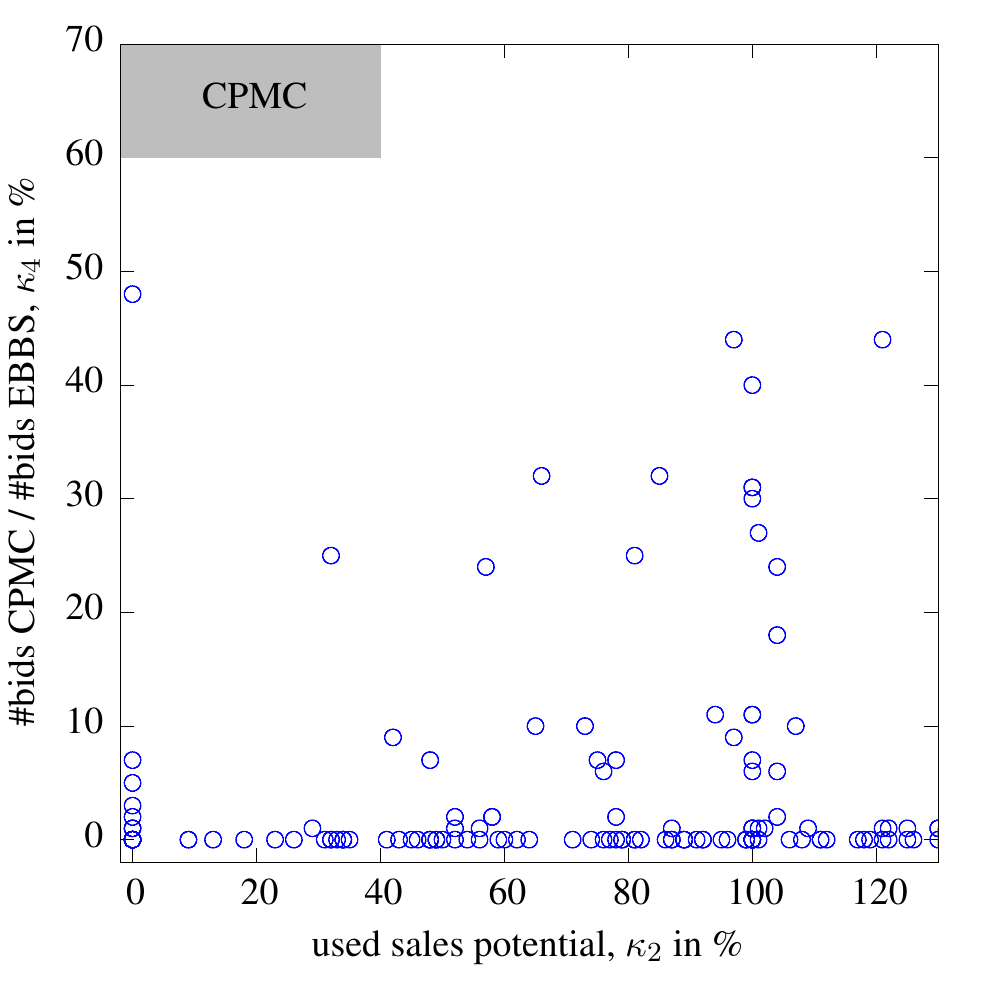}}%
\end{minipage}%
\caption{Efficiency of \psc{} and \cpmc.}%
\label{fig:efficiency}%
\end{figure}

To gain more insights on the efficiency and effectiveness of the bidding strategies look at Figure~\ref{fig:efficiency}.
On both diagrams, the abscissa shows the effectiveness measured in the used sales potential $\kappa^\phi_2$.
Higher values are better.
Note, a few data points with $\kappa^\phi_2 \geq 120$ are not shown.
The ordinate shows the efficiency measure $\kappa^\phi_4$.
Lower values are better.
Efficiency of a heuristic strategy is measured in terms of the number of computed bundle bids relative to the number of bundle bids computed by the exact strategy \ebbs.
Each circle in the diagrams shows the results for one of the 174 test instances solved by the strategy \psc{} (left diagram) and by the strategy \cpmc{} (right diagram).
Circles in lower-right area of a diagram indicate better results.

Strategy \cpmc{} is more efficient than \psc{}.
Although both strategies solve many instances with $\kappa_2 \sim 100\%$, \cpmc{} requires less bundle bids to achieve these results.
Actually, for most of the instances the number of computed bundle bids is significantly less than five percent compared to \ebbs.
In contrast, \psc{} requires for all instances significantly more than five percent of the bundle bids of \ebbs.
Averaged over all 174 instances, the strategy \psc{} requires only \emph{36.1 percent} of the bundle bids of \ebbs{} in order to achieve 90.8 percent of the sales volume won by strategy \ebbs.
The bids of strategy \cpmc{} are even more efficient.
On average, the strategy \cpmc{} requires only \emph{4.3 percent} of the bundle bids of \ebbs{} in order to achieve 80.9 percent of the sales volume won by strategy \ebbs.

\subsubsection{Effects on auctioneer's total procurement costs}

The auctioneer's total procurement costs are also affected by the choice of a bidding strategy.
Table~\ref{tab:auctioneer} shows that the maximum increase of the procurement costs is about 11 percent if the carrier uses strategy \psc.
The other strategies \cpmc, \ran, and \rann{} provide an inferior worst-case performance, because the cost increase for each strategy is about 25 percent.
For the auctioneer, strategy \psc{} is more competitive than strategy \cpmc{} which follows from the mean and median values of Table~\ref{tab:auctioneer}.
As the table indicates, the auctioneer will probably be able to significantly reduce his procurement costs, if he provides advanced data and decision support tools in order to support the bidding process of the carriers.

\begin{table}[htb]
\centering
\begin{tabular}{lrrrrllrrrr} \toprule
                    & $\kappa^{\psc}_1$ & $\kappa^{\cpmc}_1$ & $\kappa^{\ran}_1$ & $\kappa^{\rann}_1$ & \qquad &
                    & $\kappa^{\psc}_1$ & $\kappa^{\cpmc}_1$ & $\kappa^{\ran}_1$ & $\kappa^{\rann}_1$ \\ \cmidrule(lr){1-5} \cmidrule(lr){7-11}
\bf min             & 0.0  & 0.0 & 0.0 & 0.0 & & \bf 25\%-quantile & 0.0 & 1.0 & 2.9 & 2.2 \\
\bf max             & 11.1 & 23.1 & 28.6 & 22.4 & & \bf median        & 0.8 & 2.8 & 6.3 & 5.0  \\
\bf mean            & 1.4 & 3.5 & 7.8 & 6.7  & & \bf 75\%-quantile   & 2.4 & 4.9 & 11.9 & 9.7  \\ \bottomrule
\end{tabular}
\caption{Increase $\kappa^\phi_1$ of the auctioneer's total procurement costs $f^a$ over 174 test instances.} \label{tab:auctioneer}
\end{table}

\section{Conclusion and outlook} \label{sec:conclusion}
In order to support a freight carrier in a combinatorial transport auction we introduced and evaluated three strategies for truthful bundle bidding.
The bidding strategies help a bidder (i.e., a carrier) to decide on which subsets of the set of tendered requests to bid.
The bidder's goal is to maximize the sum of the prices of those bundle bids accepted by the auctioneer (i.e., the shipper).
All three strategies do not require information about the bundle bids of the rivaling carriers which is advantageous because such information is often very difficult to estimate.

The bidding strategy \ebbs{} is called \emph{exact}, because it guarantees the same results as bidding on each request combination in the powerset of the set of tendered requests.
In contrast to such a na\"{\i}ve brute-force strategy, \ebbs{} computes a significantly smaller number of bundle bids by using the notion of \emph{elementary bundle bids}.
The exact bidding strategy \ebbs{} was motivated by showing that it is---given the transport and auction scenarios at hand---necessary and sufficient to compute all elementary bundle bids in order to achieve the same outcome as a brute-force strategy.
The bidding strategies denoted as \psc{} and \cpmc{} are called \emph{heuristic}.
The idea of these heuristic strategies is to generate only those elementary bundle bids which are supersets of \emph{promising} request combinations.
Bid strategy \psc{} identifies promising request combinations by measuring pairwise synergies through modified Clarke-and-Wright savings values.
In contrast, bid strategy \cpmc{} identifies promising request combinations by solving the capacitated p-median problem.
The computational evaluation showed that \ebbs{} may be used in combinatorial transport auctions with up to 40 tendered requests.
This is about twice the size of most previous truthful bidding strategies.
On average, bid strategy \psc{} requires 36 percent of the bundle bids generated by \ebbs{} in order to achieve 91 percent of the possible revenue.
Bid strategy \cpmc{} is more efficient, it computes only 4.3 percent of the bundle bids generated by \ebbs{}, but achieves 80.9 percent of the possible revenue.
However, the results of \psc{} are more robust compared to \cpmc{}.

By showing that heuristic bundle bidding strategies are viable and proposing ways and means to evaluate the performance of (heuristic) strategies for bundle bidding a basis for future research has been created.
If we learn more about promising request combinations and heuristic bidding strategies, carriers as well as shippers may benefit.
Carriers increase their chance to win additional business.
The entrance barriers to successfully bid in transport auctions are reduced, because the computational effort to generate bids decreases with more powerful strategies.
On the one hand, future research should study features of winning bundle bids and non-winning bundle bids in order to gain more insights into characteristics of \emph{promising} request combinations.
The bidding strategies should be adapted to other transport auction scenarios with, e.g., existing transport commitments, general pickup-and-delivery requests, or requests with time windows.
On the other hand, shippers also benefit from effective bidding strategies.
Therefore, shippers should provide tools that comply with the carrier's private information requirements in order to support the carriers to generate competitive bundle bids.

\section*{Acknowledgements}
The cooperative junior research group on \emph{Computational Logistics} is funded by the University of Bremen in line with the Excellence Initiative of German federal and state governments.


\bibliographystyle{model2-names} 

\section*{Supplementary data / electronic appendix}
\begin{landscape}
\small
\begin{longtable}{rrrrrrrrrrrrrrrrrrrrr}
\caption{Detailed comparison of \ebbs, \psc, \cpmc, \ran, and \rann} \label{tab:results}\\
\toprule
\multicolumn{3}{l}{Instance} & \multicolumn{5}{l}{Auctioneer} & \multicolumn{5}{l}{Bidding carrier, used potential} & \multicolumn{5}{l}{Bidding carrier, success rate (\%)} & \multicolumn{3}{l}{Time (s)} \\
 \cmidrule(lr){1-3} \cmidrule(lr){4-8} \cmidrule(lr){9-13} \cmidrule(lr){14-18} \cmidrule(lr){19-21}
 Id & $|R|$ & $cap$ & $f^{a}(W^{\ebbs})$ & $\kappa^{\psc}_1$ & $\kappa^{\cpmc}_1$ & $\kappa^{\ran}_1$ & $\kappa^{\rann}_1$ &
 $f^{b}(W^{\ebbs})$ & $\kappa_2^{\psc}$ & $\kappa_2^{\cpmc}$ & $\kappa_2^{\ran}$ & $\kappa_2^{\rann}$ &
 $\kappa_3^{\ebbs}$ & $\kappa_3^{\psc}$ & $\kappa_3^{\cpmc}$ & $\kappa_3^{\ran}$ & \multicolumn{1}{c}{$\kappa_3^{\rann}$} &
 \multicolumn{1}{c}{\ebbs} & \multicolumn{1}{c}{\psc} & \multicolumn{1}{c}{\cpmc} \\ \midrule
\endfirsthead
\caption[]{Detailed comparison of \ebbs, \psc, \cpmc, \ran, and \rann}\\
\toprule
\multicolumn{3}{l}{Instance} & \multicolumn{5}{l}{Auctioneer} & \multicolumn{5}{l}{Bidding carrier, used potential} & \multicolumn{5}{l}{Bidding carrier, success rate (\%)} & \multicolumn{3}{l}{Time (s)} \\
 \cmidrule(lr){1-3} \cmidrule(lr){4-8} \cmidrule(lr){9-13} \cmidrule(lr){14-18} \cmidrule(lr){19-21}
 Id & $|R|$ & $cap$ & $f^{a}(W^{\ebbs})$ & $\kappa^{\psc}_1$ & $\kappa^{\cpmc}_1$ & $\kappa^{\ran}_1$ & $\kappa^{\rann}_1$ &
 $f^{b}(W^{\ebbs})$ & $\kappa_2^{\psc}$ & $\kappa_2^{\cpmc}$ & $\kappa_2^{\ran}$ & $\kappa_2^{\rann}$ &
 $\kappa_3^{\ebbs}$ & $\kappa_3^{\psc}$ & $\kappa_3^{\cpmc}$ & $\kappa_3^{\ran}$ & \multicolumn{1}{c}{$\kappa_3^{\rann}$} &
 \multicolumn{1}{c}{\ebbs} & \multicolumn{1}{c}{\psc} & \multicolumn{1}{c}{\cpmc} \\ \midrule
\endhead
1 & 15 & 50 & 295.96 & -- & -- & -- & -- & 0 & -- & -- & -- & -- & -- & -- & -- & -- & -- & 0 & 0 & 0 \\
2 & 15 & 50 & 335.17 & -- & -- & -- & -- & 0 & -- & -- & -- & -- & -- & -- & -- & -- & -- & 0 & 0 & 0 \\
3 & 15 & 50 & 297.74 & 3.0 & 0.0 & 3.0 & 3.0 & 68 & 0.0 & 100.0 & 0.0 & 0.0 & 0.1 & 0.0 & 0.2 & 0.0 & 0.0 & 0 & 0 & 0 \\
4 & 15 & 50 & 355.98 & 2.5 & 4.9 & 4.9 & 4.9 & 209 & 98.1 & 0.0 & 0.0 & 0.0 & 0.1 & 0.2 & 0.0 & 0.0 & 0.0 & 0 & 0 & 0 \\
5 & 15 & 50 & 257.60 & 0.0 & 0.0 & 1.2 & 0.0 & 79 & 100.0 & 100.0 & 0.0 & 100.0 & 0.1 & 0.3 & 1.2 & 0.0 & 2.1 & 0 & 0 & 0 \\
6 & 15 & 60 & 285.02 & -- & -- & -- & -- & 0 & -- & -- & -- & -- & -- & -- & -- & -- & -- & 0 & 0 & 0 \\
7 & 15 & 60 & 290.62 & 0.0 & 0.7 & 0.7 & 0.7 & 72 & 100.0 & 0.0 & 0.0 & 0.0 & 0.3 & 0.5 & 0.0 & 0.0 & 0.0 & 0 & 0 & 0 \\
8 & 15 & 60 & 291.64 & 0.0 & 2.4 & 5.3 & 5.3 & 79 & 100.0 & 236.7 & 0.0 & 0.0 & 0.0 & 0.2 & 3.2 & 0.0 & 0.0 & 0 & 0 & 0 \\
9 & 15 & 60 & 315.82 & -- & -- & -- & -- & 0 & -- & -- & -- & -- & -- & -- & -- & -- & -- & 0 & 0 & 0 \\
10 & 15 & 60 & 239.47 & 2.5 & 2.5 & 6.3 & 3.0 & 50 & 250.0 & 176.0 & 180.0 & 92.0 & 0.1 & 0.4 & 2.2 & 13.3 & 0.8 & 0 & 0 & 0 \\
11 & 15 & 70 & 273.01 & -- & -- & -- & -- & 0 & -- & -- & -- & -- & -- & -- & -- & -- & -- & 0 & 0 & 0 \\
12 & 15 & 70 & 291.33 & 0.0 & 0.0 & 1.0 & 1.0 & 145 & 100.0 & 100.0 & 0.0 & 0.0 & 0.3 & 0.7 & 0.9 & 0.0 & 0.0 & 0 & 0 & 0 \\
13 & 15 & 70 & 264.56 & -- & -- & -- & -- & 0 & -- & -- & -- & -- & -- & -- & -- & -- & -- & 1 & 0 & 0 \\
14 & 15 & 70 & 271.08 & 0.8 & 12.1 & 12.7 & 6.8 & 114 & 93.9 & 187.7 & 0.0 & 99.1 & 0.0 & 0.0 & 22.2 & 0.0 & 2.4 & 1 & 1 & 0 \\
15 & 15 & 70 & 214.04 & 0.0 & 3.4 & 7.0 & 6.6 & 118 & 100.0 & 130.5 & 0.0 & 77.1 & 0.1 & 0.2 & 6.9 & 0.0 & 1.9 & 0 & 0 & 0 \\
16 & 15 & 80 & 250.42 & 0.0 & 0.0 & 1.1 & 1.1 & 98 & 100.0 & 100.0 & 0.0 & 0.0 & 0.0 & 0.1 & 0.4 & 0.0 & 0.0 & 0 & 0 & 0 \\
17 & 15 & 80 & 286.39 & 0.0 & 2.9 & 2.9 & 2.9 & 55 & 278.2 & 0.0 & 0.0 & 0.0 & 0.1 & 0.5 & 0.0 & 0.0 & 0.0 & 0 & 0 & 0 \\
18 & 15 & 80 & 275.40 & 0.0 & 0.0 & 3.2 & 3.2 & 98 & 100.0 & 100.0 & 0.0 & 0.0 & 0.0 & 0.1 & 1.6 & 0.0 & 0.0 & 3 & 0 & 0 \\
19 & 15 & 80 & 261.44 & 0.0 & 0.8 & 1.9 & 1.9 & 95 & 100.0 & 104.2 & 0.0 & 0.0 & 0.0 & 0.0 & 0.5 & 0.0 & 0.0 & 4 & 3 & 0 \\
20 & 15 & 80 & 184.50 & 0.5 & 1.4 & 5.2 & 2.0 & 41 & 124.4 & 107.3 & 119.5 & 112.2 & 0.0 & 0.1 & 0.2 & 6.7 & 0.5 & 1 & 0 & 0 \\ [0.3em]
21 & 20 & 50 & 392.36 & 2.5 & 0.3 & 3.3 & 0.0 & 136 & 120.6 & 48.5 & 0.0 & 100.0 & 0.1 & 0.2 & 0.9 & 0.0 & 0.5 & 0 & 0 & 0 \\
22 & 20 & 50 & 453.24 & -- & -- & -- & -- & 0 & -- & -- & -- & -- & -- & -- & -- & -- & -- & 0 & 0 & 0 \\
23 & 20 & 50 & 385.32 & 1.8 & 0.5 & 6.5 & 6.2 & 311 & 86.2 & 100.6 & 0.0 & 22.8 & 0.1 & 0.1 & 0.2 & 0.0 & 0.2 & 1 & 0 & 0 \\
24 & 20 & 50 & 590.37 & 0.8 & 1.1 & 2.9 & 0.0 & 96 & 160.4 & 109.4 & 0.0 & 100.0 & 0.0 & 0.0 & 1.4 & 0.0 & 0.7 & 1 & 0 & 1 \\
25 & 20 & 50 & 424.78 & 1.9 & 2.7 & 2.7 & 2.7 & 209 & 89.0 & 0.0 & 0.0 & 0.0 & 0.2 & 0.4 & 0.0 & 0.0 & 0.0 & 0 & 0 & 0 \\
26 & 20 & 60 & 390.34 & 0.3 & 3.6 & 3.6 & 3.4 & 226 & 89.8 & 0.0 & 0.0 & 35.4 & 0.1 & 0.1 & 0.0 & 0.0 & 0.1 & 0 & 0 & 0 \\
27 & 20 & 60 & 419.08 & 0.0 & 0.0 & 0.0 & 0.4 & 148 & 100.0 & 100.0 & 44.6 & 0.0 & 0.2 & 0.4 & 1.6 & 5.0 & 0.0 & 0 & 0 & 0 \\
28 & 20 & 60 & 371.19 & 3.8 & 5.7 & 10.9 & 10.9 & 323 & 101.2 & 106.5 & 0.0 & 0.0 & 0.0 & 0.1 & 6.0 & 0.0 & 0.0 & 10 & 0 & 0 \\
29 & 20 & 60 & 510.38 & 0.0 & 0.0 & 7.7 & 7.7 & 269 & 100.0 & 100.0 & 0.0 & 0.0 & 0.0 & 0.0 & 4.4 & 0.0 & 0.0 & 6 & 5 & 0 \\
30 & 20 & 60 & 354.12 & -- & -- & -- & -- & 0 & -- & -- & -- & -- & -- & -- & -- & -- & -- & 1 & 0 & 1 \\
31 & 20 & 70 & 361.28 & 1.1 & 0.5 & 1.7 & 1.7 & 157 & 43.9 & 52.9 & 0.0 & 0.0 & 0.0 & 0.0 & 0.8 & 0.0 & 0.0 & 3 & 1 & 1 \\
32 & 20 & 70 & 383.11 & 0.0 & 2.8 & 2.8 & 2.8 & 119 & 100.0 & 0.0 & 0.0 & 0.0 & 0.1 & 0.2 & 0.0 & 0.0 & 0.0 & 0 & 0 & 0 \\
33 & 20 & 70 & 353.62 & 1.5 & 2.1 & 7.0 & 7.0 & 255 & 78.0 & 141.6 & 0.0 & 0.0 & 0.0 & 0.0 & 36.4 & 0.0 & 0.0 & 150 & 5 & 2 \\
34 & 20 & 70 & 440.50 & 0.7 & 4.9 & 11.5 & 11.5 & 275 & 98.5 & 60.7 & 0.0 & 0.0 & 0.0 & 0.0 & 0.6 & 0.0 & 0.0 & 121 & 33 & 1 \\
35 & 20 & 70 & 317.00 & 0.0 & 1.8 & 6.8 & 4.6 & 142 & 100.0 & 58.5 & 0.0 & 35.2 & 0.0 & 0.1 & 0.4 & 0.0 & 1.0 & 5 & 0 & 0 \\
36 & 20 & 80 & 350.96 & 0.8 & 0.0 & 3.0 & 3.0 & 211 & 70.6 & 100.0 & 0.0 & 0.0 & 0.0 & 0.0 & 1.1 & 0.0 & 0.0 & 24 & 3 & 1 \\
37 & 20 & 80 & 343.74 & 2.8 & 2.9 & 2.8 & 3.1 & 133 & 52.6 & 48.1 & 52.6 & 0.0 & 0.0 & 0.1 & 3.1 & 5.0 & 0.0 & 1 & 0 & 0 \\
38 & 20 & 80 & 329.25 & -- & -- & -- & -- & 0 & -- & -- & -- & -- & -- & -- & -- & -- & -- & 712 & 30 & 2 \\
39 & 20 & 80 & 421.99 & -- & -- & -- & -- & 0 & -- & -- & -- & -- & -- & -- & -- & -- & -- & 655 & 71 & 1 \\
40 & 20 & 80 & 295.96 & 0.7 & 3.4 & 8.1 & 8.1 & 140 & 97.9 & 125.0 & 0.0 & 0.0 & 0.0 & 0.1 & 1.1 & 0.0 & 0.0 & 48 & 2 & 1 \\  [0.3em]
41 & 25 & 50 & 512.32 & 0.0 & 5.3 & 8.2 & 6.7 & 294 & 100.0 & 49.0 & 15.0 & 21.1 & 0.1 & 0.2 & 17.6 & 4.0 & 0.1 & 2 & 0 & 1 \\
42 & 25 & 50 & 617.06 & 0.0 & 0.8 & 1.4 & 1.4 & 255 & 100.0 & 23.5 & 0.0 & 0.0 & 0.3 & 0.7 & 10.0 & 0.0 & 0.0 & 0 & 0 & 2 \\
43 & 25 & 50 & 506.16 & 1.6 & 2.2 & 9.5 & 9.9 & 450 & 68.9 & 87.3 & 13.6 & 0.0 & 0.0 & 0.1 & 19.2 & 4.0 & 0.0 & 78 & 9 & 0 \\
44 & 25 & 50 & 763.19 & 0.0 & 3.9 & 5.1 & 6.9 & 163 & 160.1 & 291.4 & 102.5 & 100.0 & 0.0 & 0.0 & 3.2 & 4.0 & 0.3 & 32 & 1 & 4 \\
45 & 25 & 50 & 481.48 & 0.1 & 1.3 & 3.5 & 0.8 & 274 & 96.4 & 78.1 & 0.0 & 96.4 & 0.1 & 0.1 & 3.1 & 0.0 & 1.9 & 1 & 0 & 3 \\
46 & 25 & 60 & 472.99 & 0.3 & 2.2 & 8.6 & 8.2 & 321 & 132.7 & 121.2 & 0.0 & 48.0 & 0.0 & 0.1 & 1.8 & 0.0 & 0.1 & 30 & 2 & 0 \\
47 & 25 & 60 & 550.27 & 0.7 & 3.4 & 6.7 & 4.0 & 310 & 121.6 & 122.6 & 0.0 & 24.8 & 0.1 & 0.4 & 10.9 & 0.0 & 0.3 & 0 & 0 & 3 \\
48 & 25 & 60 & 472.72 & 2.0 & 4.1 & 8.8 & 8.8 & 349 & 77.1 & 141.0 & 0.0 & 0.0 & 0.0 & 0.0 & 9.2 & 0.0 & 0.0 & 760 & 5 & 2 \\
49 & 25 & 60 & 662.70 & 0.0 & 4.4 & 8.9 & 8.3 & 270 & 100.0 & 77.8 & 0.0 & 41.5 & 0.0 & 0.0 & 3.0 & 0.0 & 0.4 & 472 & 2 & 1 \\
50 & 25 & 60 & 425.41 & 3.4 & 3.5 & 10.6 & 5.8 & 192 & 84.9 & 79.2 & 15.6 & 89.6 & 0.0 & 0.0 & 10.0 & 4.0 & 0.5 & 12 & 2 & 2 \\
51 & 25 & 70 & 440.18 & 1.6 & 5.2 & 12.4 & 9.4 & 382 & 85.3 & 96.3 & 0.0 & 19.1 & 0.0 & 0.0 & 2.0 & 0.0 & 0.1 & 224 & 25 & 3 \\
52 & 25 & 70 & 514.24 & 0.0 & 2.1 & 5.1 & 3.7 & 299 & 100.0 & 87.6 & 0.0 & 46.2 & 0.0 & 0.2 & 3.2 & 0.0 & 0.7 & 5 & 0 & 1 \\
53 & 25 & 70 & 448.90 & 2.9 & 9.3 & 19.3 & 19.3 & 374 & 103.5 & 81.6 & 0.0 & 52.4 & 0.0 & 0.0 & 0.0 & 0.0 & 0.4 & 4,954 & 70 & 523 \\
54 & 25 & 70 & 579.54 & 4.9 & 4.9 & 15.3 & 15.3 & 277 & 59.6 & 59.6 & 0.0 & 0.0 & 0.0 & 0.0 & 1.2 & 0.0 & 0.0 & 3,595 & 5 & 2 \\
55 & 25 & 70 & 371.72 & 4.1 & 2.4 & 14.8 & 10.4 & 353 & 36.3 & 9.6 & 0.0 & 24.1 & 0.0 & 0.0 & 1.1 & 0.0 & 0.5 & 43 & 20 & 2 \\
56 & 25 & 80 & 402.82 & 4.2 & 9.7 & 17.1 & 15.8 & 338 & 105.0 & 130.8 & 0.0 & 21.6 & 0.0 & 0.0 & 5.4 & 0.0 & 0.1 & 1,224 & 231 & 1 \\
57 & 25 & 80 & 477.05 & 3.0 & 1.6 & 9.7 & 8.6 & 401 & 87.5 & 101.5 & 0.0 & 39.7 & 0.0 & 0.1 & 5.1 & 0.0 & 0.8 & 26 & 1 & 1 \\
58 & 25 & 80 & 419.36 & 4.5 & 5.4 & 21.4 & 17.9 & 392 & 86.7 & 71.4 & 0.0 & 41.6 & 0.0 & 0.0 & 0.3 & 0.0 & 0.1 & 27,705 & 537 & 21 \\
59 & 25 & 80 & 516.32 & 9.0 & 4.2 & 17.1 & 17.1 & 392 & 0.0 & 137.2 & 0.0 & 0.0 & 0.0 & 0.0 & 100.0 & 0.0 & 0.0 & 23,597 & 92 & 2 \\
60 & 25 & 80 & 325.34 & 4.5 & 5.7 & 15.1 & 14.9 & 214 & 44.9 & 136.4 & 0.0 & 43.0 & 0.0 & 0.0 & 0.1 & 0.0 & 0.2 & 789 & 160 & 4 \\ [0.3em]
61 & 30 & 50 & 594.02 & 1.2 & 9.0 & 13.3 & 7.5 & 419 & 114.6 & 131.3 & 0.0 & 54.2 & 0.0 & 0.1 & 53.3 & 0.0 & 0.1 & 42 & 4 & 6 \\
62 & 30 & 50 & 686.54 & 0.0 & 1.8 & 1.8 & 1.3 & 158 & 100.0 & 0.0 & 0.0 & 30.4 & 0.2 & 0.3 & 0.0 & 0.0 & 0.5 & 1 & 0 & 3 \\
63 & 30 & 50 & 577.94 & 0.7 & 2.5 & 15.7 & 11.5 & 434 & 96.1 & 108.8 & 0.0 & 37.6 & 0.0 & 0.0 & 6.5 & 0.0 & 0.2 & 332 & 32 & 1 \\
64 & 30 & 50 & 951.51 & 4.3 & 7.8 & 12.0 & 9.8 & 542 & 60.5 & 160.7 & 30.1 & 70.7 & 0.0 & 0.0 & 3.4 & 3.3 & 0.7 & 735 & 16 & 4 \\
65 & 30 & 50 & 547.29 & 0.2 & 3.5 & 11.1 & 6.7 & 357 & 87.1 & 112.6 & 26.9 & 47.3 & 0.0 & 0.1 & 14.3 & 3.3 & 0.7 & 23 & 2 & 5 \\
66 & 30 & 60 & 543.58 & 0.6 & 0.7 & 9.8 & 6.5 & 496 & 66.5 & 100.8 & 0.0 & 13.3 & 0.0 & 0.0 & 1.3 & 0.0 & 0.0 & 558 & 82 & 5 \\
67 & 30 & 60 & 630.46 & 1.8 & 6.4 & 7.9 & 6.5 & 415 & 104.1 & 32.3 & 0.0 & 51.6 & 0.1 & 0.3 & 20.0 & 0.0 & 1.0 & 1 & 0 & 3 \\
68 & 30 & 60 & 526.40 & 2.1 & 2.7 & 22.3 & 21.3 & 504 & 74.8 & 97.8 & 0.0 & 19.2 & 0.0 & 0.0 & 0.0 & 0.0 & 0.1 & 3,076 & 92 & 884 \\
69 & 30 & 60 & 782.77 & 11.1 & 11.2 & 16.4 & 14.3 & 488 & 34.4 & 78.3 & 0.0 & 35.2 & 0.0 & 0.0 & 1.8 & 0.0 & 0.3 & 6,565 & 335 & 3 \\
70 & 30 & 60 & 482.31 & 1.5 & 6.4 & 14.3 & 11.2 & 244 & 99.6 & 198.0 & 14.8 & 108.2 & 0.0 & 0.0 & 21.6 & 3.3 & 0.2 & 386 & 64 & 3 \\
71 & 30 & 70 & 496.60 & 2.4 & 11.9 & 17.8 & 16.9 & 448 & 87.5 & 95.8 & 0.0 & 25.2 & 0.0 & 0.0 & 0.4 & 0.0 & 0.0 & 3,214 & 210 & 4 \\
72 & 30 & 70 & 569.62 & 0.6 & 2.0 & 11.2 & 11.4 & 418 & 107.4 & 112.0 & 20.8 & 35.4 & 0.0 & 0.1 & 43.8 & 3.3 & 0.4 & 29 & 1 & 3 \\
73 & 30 & 70 & 492.40 & 2.6 & 9.3 & 18.1 & 18.1 & 389 & 88.9 & 138.3 & 0.0 & 0.0 & 0.0 & 0.0 & 60.0 & 0.0 & 0.0 & 23,311 & 722 & 3 \\
74 & 30 & 70 & 715.25 & 5.9 & 5.0 & 13.0 & 11.4 & 289 & 67.1 & 191.3 & 0.0 & 67.8 & 0.0 & 0.0 & 0.6 & 0.0 & 0.1 & 65,971 & 7,823 & 11 \\
75 & 30 & 70 & 414.17 & 1.7 & 4.4 & 19.1 & 14.0 & 257 & 106.6 & 126.8 & 0.0 & 105.4 & 0.0 & 0.0 & 2.5 & 0.0 & 0.2 & 2,267 & 1,044 & 3 \\
76 & 30 & 80 & 459.84 & 4.0 & 8.1 & 20.3 & 20.3 & 448 & 101.6 & 104.5 & 0.0 & 0.0 & 0.0 & 0.0 & 0.0 & 0.0 & 0.0 & 21,774 & 1,220 & 329 \\
77 & 30 & 80 & 526.80 & 1.7 & 6.3 & 16.8 & 18.8 & 450 & 87.3 & 104.7 & 18.2 & 0.0 & 0.0 & 0.0 & 0.1 & 3.3 & 0.0 & 295 & 11 & 9 \\
80 & 30 & 80 & 380.93 & 5.2 & 4.6 & 18.9 & 16.4 & 287 & 109.8 & 119.5 & 30.7 & 48.1 & 0.0 & 0.0 & 0.8 & 3.3 & 0.3 & 12,712 & 4,360 & 1 \\ [0.3em]
81 & 35 & 50 & 719.04 & 0.0 & 5.0 & 13.6 & 8.9 & 584 & 100.5 & 85.1 & 0.0 & 50.3 & 0.0 & 0.1 & 0.1 & 0.0 & 0.1 & 55 & 5 & 3 \\
82 & 35 & 50 & 821.34 & 0.0 & 1.1 & 6.8 & 4.3 & 430 & 100.0 & 121.6 & 14.0 & 47.0 & 0.2 & 0.5 & 40.0 & 2.9 & 0.6 & 1 & 0 & 3 \\
85 & 35 & 50 & 719.87 & 0.6 & 3.0 & 11.3 & 9.0 & 511 & 88.3 & 93.0 & 17.4 & 43.1 & 0.0 & 0.1 & 4.5 & 2.9 & 0.2 & 63 & 12 & 5 \\
87 & 35 & 60 & 719.88 & 0.0 & 2.2 & 7.5 & 7.0 & 462 & 100.0 & 73.4 & 0.0 & 14.9 & 0.1 & 0.2 & 0.5 & 0.0 & 0.1 & 5 & 0 & 3 \\
91 & 35 & 70 & 582.76 & 3.5 & 6.3 & 21.1 & 18.8 & 527 & 109.7 & 86.5 & 0.0 & 13.7 & 0.0 & 0.0 & 1.7 & 0.0 & 0.1 & 14,213 & 683 & 3 \\
92 & 35 & 70 & 663.64 & 1.6 & 11.0 & 11.9 & 11.6 & 626 & 80.4 & 13.1 & 12.0 & 10.2 & 0.0 & 0.1 & 10.0 & 2.9 & 0.3 & 78 & 3 & 3 \\
93 & 35 & 70 & 593.25 & 2.5 & 5.8 & 17.4 & 17.2 & 571 & 78.3 & 89.3 & 0.0 & 15.9 & 0.0 & 0.0 & 0.4 & 0.0 & 0.1 & 88,809 & 2,838 & 30 \\
95 & 35 & 70 & 549.64 & 2.1 & 2.4 & 14.4 & 12.5 & 430 & 110.0 & 99.5 & 0.0 & 19.3 & 0.0 & 0.0 & 2.3 & 0.0 & 0.1 & 6,206 & 2,342 & 3 \\
96 & 35 & 80 & 559.28 & 1.6 & 7.1 & 23.2 & 16.2 & 506 & 99.4 & 118.4 & 0.0 & 25.9 & 0.0 & 0.0 & 3.0 & 0.0 & 0.0 & 72,865 & 5,623 & 5 \\
97 & 35 & 80 & 612.58 & 1.1 & 8.2 & 15.3 & 13.5 & 410 & 92.0 & 91.5 & 0.0 & 25.1 & 0.0 & 0.0 & 38.5 & 0.0 & 0.2 & 1,140 & 43 & 3 \\
100 & 35 & 80 & 497.49 & 2.2 & 4.9 & 16.1 & 16.1 & 283 & 132.9 & 134.6 & 0.0 & 0.0 & 0.0 & 0.0 & 27.3 & 0.0 & 0.0 & 35,111 & 13,440 & 3 \\ [0.3em]
101 & 40 & 50 & 822.58 & 0.4 & 4.9 & 12.1 & 7.1 & 695 & 83.2 & 82.6 & 0.0 & 33.7 & 0.0 & 0.0 & 29.2 & 0.0 & 0.0 & 761 & 106 & 3 \\
102 & 40 & 50 & 929.92 & 0.0 & 5.6 & 10.1 & 2.8 & 743 & 100.0 & 78.9 & 6.7 & 46.2 & 0.3 & 0.6 & 3.2 & 2.5 & 0.4 & 1 & 0 & 4 \\
103 & 40 & 50 & 833.67 & 0.0 & 2.3 & 18.0 & 16.6 & 595 & 102.2 & 136.0 & 25.5 & 18.5 & 0.0 & 0.0 & 5.9 & 5.0 & 0.0 & 5,307 & 1,141 & 3 \\
104 & 40 & 50 & 1,282.77 & 5.7 & 2.4 & 17.1 & 13.4 & 839 & 59.2 & 153.6 & 13.3 & 49.2 & 0.0 & 0.0 & 0.0 & 2.5 & 0.1 & 64,490 & 32,635 & 2,050 \\
105 & 40 & 50 & 861.04 & 0.8 & 5.6 & 12.8 & 9.2 & 434 & 97.9 & 149.1 & 18.4 & 68.2 & 0.0 & 0.0 & 21.6 & 2.5 & 0.1 & 174 & 54 & 5 \\
106 & 40 & 60 & 740.78 & 1.6 & 10.4 & 21.9 & 17.2 & 644 & 98.0 & 81.4 & 0.0 & 28.1 & 0.0 & 0.0 & 5.6 & 0.0 & 0.1 & 11,224 & 1,250 & 3 \\
107 & 40 & 60 & 837.59 & 1.0 & 4.0 & 13.1 & 11.5 & 686 & 87.3 & 97.4 & 24.9 & 46.5 & 0.1 & 0.2 & 0.7 & 5.0 & 0.6 & 28 & 1 & 3 \\
108 & 40 & 60 & 728.93 & 2.7 & 4.7 & 23.1 & 18.2 & 476 & 98.7 & 133.4 & 22.9 & 49.8 & 0.0 & 0.0 & 3.3 & 2.5 & 0.0 & 66,825 & 4,622 & 6 \\
110 & 40 & 60 & 762.91 & 0.8 & 5.4 & 18.1 & 13.3 & 566 & 82.2 & 111.5 & 0.0 & 71.0 & 0.0 & 0.0 & 3.3 & 0.0 & 0.1 & 3,012 & 870 & 3 \\
111 & 40 & 70 & 667.06 & 3.5 & 23.1 & 28.6 & 22.4 & 594 & 113.6 & 46.3 & 0.0 & 37.5 & 0.0 & 0.0 & 33.3 & 0.0 & 0.1 & 116,071 & 11,161 & 3 \\
112 & 40 & 70 & 739.05 & 2.6 & 6.2 & 17.6 & 17.5 & 582 & 102.7 & 101.9 & 0.0 & 13.6 & 0.0 & 0.1 & 1.5 & 0.0 & 0.2 & 510 & 31 & 1 \\
115 & 40 & 70 & 668.00 & 2.1 & 4.4 & 21.2 & 20.5 & 647 & 96.6 & 92.1 & 0.0 & 27.2 & 0.0 & 0.0 & 3.9 & 0.0 & 0.1 & 24,365 & 7,120 & 4 \\
117 & 40 & 80 & 692.38 & 1.7 & 18.5 & 19.1 & 14.3 & 532 & 101.3 & 9.4 & 0.0 & 62.6 & 0.0 & 0.0 & 10.0 & 0.0 & 0.5 & 4,228 & 329 & 3 \\ [0.3em]
121 & 15 & 50 & 274.14 & -- & -- & -- & -- & 0 & -- & -- & -- & -- & -- & -- & -- & -- & -- & 0 & 0 & 0 \\
122 & 15 & 50 & 309.93 & 0.0 & 0.1 & 0.1 & 0.0 & 67 & 100.0 & 0.0 & 0.0 & 100.0 & 0.5 & 0.8 & 0.0 & 0.0 & 1.5 & 0 & 0 & 1 \\
123 & 15 & 50 & 283.73 & -- & -- & -- & -- & 0 & -- & -- & -- & -- & -- & -- & -- & -- & -- & 0 & 0 & 0 \\
124 & 15 & 50 & 360.05 & -- & -- & -- & -- & 0 & -- & -- & -- & -- & -- & -- & -- & -- & -- & 0 & 0 & 0 \\
125 & 15 & 50 & 256.74 & 0.0 & 0.2 & 2.7 & 0.8 & 39 & 100.0 & 94.9 & 0.0 & 112.8 & 0.1 & 0.3 & 1.2 & 0.0 & 1.6 & 0 & 0 & 2 \\
126 & 15 & 60 & 261.21 & -- & -- & -- & -- & 0 & -- & -- & -- & -- & -- & -- & -- & -- & -- & 0 & 0 & 2 \\
127 & 15 & 60 & 277.12 & -- & -- & -- & -- & 0 & -- & -- & -- & -- & -- & -- & -- & -- & -- & 0 & 0 & 1 \\
128 & 15 & 60 & 245.90 & -- & -- & -- & -- & 0 & -- & -- & -- & -- & -- & -- & -- & -- & -- & 0 & 0 & 0 \\
129 & 15 & 60 & 303.87 & -- & -- & -- & -- & 0 & -- & -- & -- & -- & -- & -- & -- & -- & -- & 0 & 0 & 2 \\
130 & 15 & 60 & 221.05 & 0.5 & 1.6 & 4.9 & 0.0 & 45 & 100.0 & 100.0 & 0.0 & 100.0 & 0.1 & 0.2 & 1.1 & 0.0 & 0.7 & 0 & 0 & 0 \\
131 & 15 & 70 & 239.67 & -- & -- & -- & -- & 0 & -- & -- & -- & -- & -- & -- & -- & -- & -- & 0 & 0 & 2 \\
132 & 15 & 70 & 264.26 & -- & -- & -- & -- & 0 & -- & -- & -- & -- & -- & -- & -- & -- & -- & 0 & 0 & 1 \\
133 & 15 & 70 & 263.42 & -- & -- & -- & -- & 0 & -- & -- & -- & -- & -- & -- & -- & -- & -- & 1 & 0 & 1 \\
134 & 15 & 70 & 262.36 & -- & -- & -- & -- & 0 & -- & -- & -- & -- & -- & -- & -- & -- & -- & 2 & 1 & 0 \\
135 & 15 & 70 & 200.47 & 2.7 & 3.5 & 3.5 & 2.0 & 45 & 113.3 & 0.0 & 0.0 & 108.9 & 0.0 & 0.1 & 0.0 & 0.0 & 1.2 & 0 & 0 & 0 \\
136 & 15 & 80 & 241.28 & -- & -- & -- & -- & 0 & -- & -- & -- & -- & -- & -- & -- & -- & -- & 0 & 0 & 1 \\
137 & 15 & 80 & 241.20 & -- & -- & -- & -- & 0 & -- & -- & -- & -- & -- & -- & -- & -- & -- & 0 & 0 & 1 \\
138 & 15 & 80 & 251.18 & -- & -- & -- & -- & 0 & -- & -- & -- & -- & -- & -- & -- & -- & -- & 5 & 0 & 1 \\
139 & 15 & 80 & 232.55 & -- & -- & -- & -- & 0 & -- & -- & -- & -- & -- & -- & -- & -- & -- & 8 & 4 & 0 \\
140 & 15 & 80 & 178.57 & -- & -- & -- & -- & 0 & -- & -- & -- & -- & -- & -- & -- & -- & -- & 1 & 0 & 1 \\ [0.3em]
141 & 20 & 50 & 373.97 & 0.0 & 0.1 & 0.1 & 0.0 & 82 & 100.0 & 0.0 & 0.0 & 100.0 & 0.1 & 0.1 & 0.0 & 0.0 & 0.2 & 0 & 0 & 1 \\
142 & 20 & 50 & 408.76 & -- & -- & -- & -- & 0 & -- & -- & -- & -- & -- & -- & -- & -- & -- & 0 & 0 & 1 \\
143 & 20 & 50 & 369.26 & 0.3 & 0.0 & 0.3 & 0.0 & 61 & 0.0 & 100.0 & 0.0 & 100.0 & 0.0 & 0.0 & 0.1 & 0.0 & 0.3 & 1 & 0 & 0 \\
144 & 20 & 50 & 561.34 & 0.0 & 0.3 & 0.3 & 0.3 & 93 & 100.0 & 0.0 & 0.0 & 0.0 & 0.0 & 0.0 & 0.0 & 0.0 & 0.0 & 1 & 0 & 3 \\
145 & 20 & 50 & 407.61 & -- & -- & -- & -- & 0 & -- & -- & -- & -- & -- & -- & -- & -- & -- & 0 & 0 & 1 \\
146 & 20 & 60 & 348.35 & -- & -- & -- & -- & 0 & -- & -- & -- & -- & -- & -- & -- & -- & -- & 0 & 0 & 1 \\
147 & 20 & 60 & 390.26 & -- & -- & -- & -- & 0 & -- & -- & -- & -- & -- & -- & -- & -- & -- & 0 & 0 & 1 \\
148 & 20 & 60 & 357.60 & 1.5 & 0.0 & 1.9 & 1.9 & 99 & 75.8 & 100.0 & 0.0 & 0.0 & 0.0 & 0.0 & 1.5 & 0.0 & 0.0 & 26 & 1 & 0 \\
149 & 20 & 60 & 477.12 & 0.0 & 0.0 & 1.0 & 1.0 & 157 & 100.0 & 100.0 & 0.0 & 0.0 & 0.0 & 0.0 & 2.2 & 0.0 & 0.0 & 19 & 7 & 2 \\
150 & 20 & 60 & 335.96 & 0.4 & 0.4 & 0.4 & 0.4 & 45 & 0.0 & 0.0 & 0.0 & 0.0 & 0.0 & 0.0 & 0.0 & 0.0 & 0.0 & 1 & 0 & 2 \\
151 & 20 & 70 & 333.91 & -- & -- & -- & -- & 0 & -- & -- & -- & -- & -- & -- & -- & -- & -- & 4 & 1 & 3 \\
152 & 20 & 70 & 341.04 & -- & -- & -- & -- & 0 & -- & -- & -- & -- & -- & -- & -- & -- & -- & 0 & 0 & 0 \\
153 & 20 & 70 & 328.78 & 3.0 & 3.0 & 3.0 & 3.0 & 107 & 0.0 & 0.0 & 0.0 & 0.0 & 0.0 & 0.0 & 0.0 & 0.0 & 0.0 & 154 & 4 & 1 \\
154 & 20 & 70 & 406.86 & -- & -- & -- & -- & 0 & -- & -- & -- & -- & -- & -- & -- & -- & -- & 136 & 32 & 1 \\
155 & 20 & 70 & 305.93 & 0.0 & 0.5 & 2.2 & 2.2 & 85 & 100.0 & 52.9 & 0.0 & 0.0 & 0.0 & 0.1 & 0.4 & 0.0 & 0.0 & 7 & 0 & 1 \\
156 & 20 & 80 & 308.28 & -- & -- & -- & -- & 0 & -- & -- & -- & -- & -- & -- & -- & -- & -- & 31 & 5 & 2 \\
157 & 20 & 80 & 345.90 & 0.0 & 0.0 & 1.2 & 1.2 & 96 & 100.0 & 100.0 & 0.0 & 0.0 & 0.0 & 0.1 & 3.1 & 0.0 & 0.0 & 1 & 0 & 0 \\
158 & 20 & 80 & 321.80 & 1.5 & 1.4 & 1.5 & 1.5 & 93 & 89.2 & 117.2 & 0.0 & 0.0 & 0.0 & 0.0 & 0.3 & 0.0 & 0.0 & 689 & 34 & 2 \\
159 & 20 & 80 & 399.66 & 0.0 & 0.0 & 2.0 & 2.0 & 164 & 100.0 & 100.0 & 0.0 & 0.0 & 0.0 & 0.0 & 0.4 & 0.0 & 0.0 & 619 & 124 & 1 \\
160 & 20 & 80 & 281.35 & 2.7 & 1.9 & 4.4 & 3.4 & 130 & 67.7 & 102.3 & 68.5 & 78.5 & 0.0 & 0.0 & 1.1 & 5.0 & 0.5 & 53 & 1 & 0 \\ [0.3em]
161 & 25 & 50 & 479.78 & -- & -- & -- & -- & 0 & -- & -- & -- & -- & -- & -- & -- & -- & -- & 1 & 0 & 0 \\
162 & 25 & 50 & 553.12 & 0.0 & 0.5 & 0.5 & 0.0 & 77 & 100.0 & 0.0 & 0.0 & 100.0 & 0.1 & 0.2 & 0.0 & 0.0 & 0.3 & 0 & 0 & 2 \\
163 & 25 & 50 & 486.77 & 0.0 & 0.0 & 5.4 & 5.4 & 154 & 100.0 & 100.0 & 0.0 & 0.0 & 0.0 & 0.0 & 7.7 & 0.0 & 0.0 & 80 & 7 & 0 \\
164 & 25 & 50 & 741.27 & 0.0 & 0.7 & 1.9 & 1.9 & 202 & 130.7 & 49.5 & 0.0 & 0.0 & 0.0 & 0.0 & 1.1 & 0.0 & 0.0 & 38 & 1 & 3 \\
165 & 25 & 50 & 464.96 & 0.5 & 0.9 & 1.6 & 1.6 & 136 & 106.6 & 58.1 & 0.0 & 0.0 & 0.1 & 0.1 & 1.0 & 0.0 & 0.0 & 1 & 0 & 2 \\
166 & 25 & 60 & 440.15 & 0.0 & 0.0 & 0.3 & 0.3 & 85 & 100.0 & 100.0 & 0.0 & 0.0 & 0.0 & 0.0 & 0.4 & 0.0 & 0.0 & 32 & 2 & 1 \\
167 & 25 & 60 & 516.60 & 0.0 & 1.4 & 2.4 & 2.4 & 257 & 100.0 & 29.6 & 0.0 & 0.0 & 0.1 & 0.2 & 2.2 & 0.0 & 0.0 & 0 & 0 & 2 \\
168 & 25 & 60 & 437.64 & 3.4 & 4.3 & 4.3 & 4.3 & 183 & 96.7 & 0.0 & 0.0 & 0.0 & 0.0 & 0.0 & 0.0 & 0.0 & 0.0 & 735 & 9 & 1 \\
169 & 25 & 60 & 626.35 & 0.0 & 1.6 & 1.6 & 1.6 & 162 & 100.0 & 0.0 & 0.0 & 0.0 & 0.0 & 0.0 & 0.0 & 0.0 & 0.0 & 458 & 2 & 2 \\
170 & 25 & 60 & 396.80 & 1.0 & 0.5 & 3.2 & 1.4 & 77 & 58.4 & 41.6 & 0.0 & 55.8 & 0.0 & 0.0 & 3.3 & 0.0 & 0.1 & 20 & 3 & 2 \\
171 & 25 & 70 & 421.64 & 0.0 & 0.8 & 1.0 & 1.0 & 168 & 100.0 & 32.7 & 0.0 & 0.0 & 0.0 & 0.0 & 0.4 & 0.0 & 0.0 & 216 & 35 & 3 \\
172 & 25 & 70 & 466.35 & 0.0 & 0.1 & 0.1 & 0.1 & 87 & 100.0 & 0.0 & 0.0 & 0.0 & 0.0 & 0.0 & 0.0 & 0.0 & 0.0 & 8 & 0 & 1 \\
173 & 25 & 70 & 434.06 & 0.0 & 4.5 & 6.2 & 6.2 & 258 & 100.0 & 32.2 & 0.0 & 0.0 & 0.0 & 0.0 & 0.0 & 0.0 & 0.0 & 5,149 & 92 & 608 \\
174 & 25 & 70 & 557.85 & 1.8 & 2.5 & 5.4 & 5.4 & 270 & 58.9 & 100.0 & 0.0 & 0.0 & 0.0 & 0.0 & 2.3 & 0.0 & 0.0 & 2,490 & 8 & 1 \\
175 & 25 & 70 & 348.74 & 0.0 & 2.3 & 4.0 & 4.0 & 45 & 100.0 & 195.6 & 0.0 & 0.0 & 0.0 & 0.0 & 2.2 & 0.0 & 0.0 & 142 & 34 & 1 \\
176 & 25 & 80 & 395.48 & 0.0 & 0.3 & 0.4 & 0.4 & 159 & 100.0 & 49.7 & 0.0 & 0.0 & 0.0 & 0.0 & 0.9 & 0.0 & 0.0 & 1,135 & 233 & 1 \\
177 & 25 & 80 & 452.25 & 0.0 & 0.8 & 0.8 & 0.8 & 61 & 100.0 & 0.0 & 0.0 & 0.0 & 0.0 & 0.0 & 0.0 & 0.0 & 0.0 & 56 & 1 & 2 \\
178 & 25 & 80 & 417.70 & 1.3 & 2.4 & 4.6 & 4.6 & 299 & 74.2 & 64.5 & 0.0 & 0.0 & 0.0 & 0.0 & 0.2 & 0.0 & 0.0 & 24,233 & 541 & 18 \\
179 & 25 & 80 & 504.00 & 3.9 & 3.9 & 7.5 & 7.5 & 385 & 43.1 & 43.1 & 0.0 & 0.0 & 0.0 & 0.0 & 25.0 & 0.0 & 0.0 & 26,392 & 90 & 2 \\
180 & 25 & 80 & 327.76 & 1.2 & 1.6 & 4.5 & 4.5 & 123 & 74.0 & 146.3 & 0.0 & 0.0 & 0.0 & 0.0 & 0.1 & 0.0 & 0.0 & 857 & 180 & 6 \\ [0.3em]
181 & 30 & 50 & 588.74 & 0.4 & 1.9 & 2.3 & 1.1 & 340 & 50.9 & 33.2 & 0.0 & 27.4 & 0.0 & 0.0 & 13.3 & 0.0 & 0.1 & 50 & 4 & 4 \\
182 & 30 & 50 & 635.93 & -- & -- & -- & -- & 0 & -- & -- & -- & -- & -- & -- & -- & -- & -- & 0 & 0 & 4 \\
183 & 30 & 50 & 546.94 & 0.0 & 0.1 & 2.7 & 0.1 & 154 & 100.0 & 48.7 & 0.0 & 48.7 & 0.0 & 0.0 & 1.1 & 0.0 & 0.0 & 355 & 35 & 0 \\
184 & 30 & 50 & 900.63 & 1.4 & 3.0 & 3.0 & 3.0 & 275 & 68.0 & 0.0 & 0.0 & 0.0 & 0.0 & 0.0 & 0.0 & 0.0 & 0.0 & 528 & 26 & 5 \\
185 & 30 & 50 & 518.32 & 2.4 & 3.8 & 4.8 & 3.1 & 162 & 150.0 & 26.5 & 0.0 & 23.5 & 0.0 & 0.1 & 1.8 & 0.0 & 0.0 & 39 & 6 & 5 \\
186 & 30 & 60 & 529.49 & 0.6 & 1.9 & 2.2 & 2.2 & 329 & 20.1 & 130.7 & 0.0 & 0.0 & 0.0 & 0.0 & 1.1 & 0.0 & 0.0 & 469 & 99 & 4 \\
187 & 30 & 60 & 585.66 & -0.0 & 1.8 & 2.8 & 2.8 & 136 & 100.0 & 50.7 & 0.0 & 0.0 & 0.0 & 0.1 & 10.0 & 0.0 & 0.0 & 2 & 0 & 3 \\
188 & 30 & 60 & 512.68 & 1.9 & 1.3 & 7.2 & 6.8 & 349 & 80.8 & 121.2 & 0.0 & 24.6 & 0.0 & 0.0 & 0.0 & 0.0 & 0.2 & 4,141 & 106 & 662 \\
189 & 30 & 60 & 773.85 & 2.7 & 2.7 & 3.9 & 3.9 & 288 & 64.6 & 34.4 & 0.0 & 0.0 & 0.0 & 0.0 & 0.6 & 0.0 & 0.0 & 7,920 & 400 & 3 \\
190 & 30 & 60 & 454.22 & 0.7 & 2.9 & 4.2 & 4.2 & 123 & 36.6 & 100.8 & 0.0 & 0.0 & 0.0 & 0.0 & 8.1 & 0.0 & 0.0 & 447 & 136 & 5 \\
191 & 30 & 70 & 466.06 & 2.6 & 5.2 & 5.2 & 5.2 & 233 & 85.8 & 0.0 & 0.0 & 0.0 & 0.0 & 0.0 & 0.0 & 0.0 & 0.0 & 5,216 & 233 & 3 \\
192 & 30 & 70 & 538.09 & 0.0 & 0.7 & 4.4 & 4.3 & 220 & 100.0 & 62.3 & 0.0 & 27.3 & 0.0 & 0.1 & 12.5 & 0.0 & 0.2 & 46 & 2 & 3 \\
193 & 30 & 70 & 479.64 & 2.3 & 6.2 & 6.2 & 6.2 & 275 & 88.7 & 0.0 & 0.0 & 0.0 & 0.0 & 0.0 & 0.0 & 0.0 & 0.0 & 36,019 & 773 & 3 \\
194 & 30 & 70 & 676.79 & 3.5 & 8.7 & 8.7 & 8.7 & 99 & 0.0 & 173.7 & 0.0 & 0.0 & 0.0 & 0.0 & 0.2 & 0.0 & 0.0 & 66,234 & 3,338 & 5 \\
195 & 30 & 70 & 393.02 & 2.7 & 4.3 & 4.3 & 4.3 & 133 & 66.9 & 0.0 & 0.0 & 0.0 & 0.0 & 0.0 & 0.0 & 0.0 & 0.0 & 3,528 & 915 & 5 \\
196 & 30 & 80 & 453.82 & 0.9 & 5.0 & 7.6 & 7.2 & 356 & 97.2 & 76.1 & 0.0 & 19.1 & 0.0 & 0.0 & 0.0 & 0.0 & 0.1 & 24,388 & 1,652 & 327 \\
197 & 30 & 80 & 516.45 & 0.1 & 0.2 & 3.2 & 3.2 & 278 & 54.7 & 57.6 & 0.0 & 0.0 & 0.0 & 0.0 & 0.0 & 0.0 & 0.0 & 366 & 26 & 22 \\
198 & 30 & 80 & 452.93 & 2.9 & 3.4 & 11.0 & 11.0 & 277 & 97.8 & 76.5 & 0.0 & 0.0 & 0.0 & 0.0 & 0.1 & 0.0 & 0.0 & 138,571 & 3,949 & 230 \\
200 & 30 & 80 & 360.28 & 4.3 & 6.9 & 8.9 & 8.9 & 206 & 70.9 & 87.4 & 0.0 & 41.7 & 0.0 & 0.0 & 0.4 & 0.0 & 0.1 & 17,276 & 4,349 & 1 \\ [0.3em]
201 & 35 & 50 & 696.96 & 0.0 & 0.6 & 3.2 & 3.1 & 309 & 100.0 & 66.7 & 0.0 & 29.8 & 0.0 & 0.0 & 0.0 & 0.0 & 0.0 & 80 & 19 & 15 \\
202 & 35 & 50 & 757.20 & 0.0 & 0.4 & 0.4 & 1.1 & 122 & 100.0 & 49.2 & 49.2 & 0.0 & 0.1 & 0.1 & 5.0 & 2.9 & 0.0 & 0 & 0 & 4 \\
203 & 35 & 50 & 674.83 & 0.0 & 2.6 & 6.9 & 6.9 & 397 & 100.0 & 81.9 & 0.0 & 0.0 & 0.0 & 0.0 & 30.8 & 0.0 & 0.0 & 1,201 & 155 & 3 \\
204 & 35 & 50 & 966.87 & 1.0 & 4.9 & 7.8 & 7.7 & 289 & 66.8 & 34.6 & 0.0 & 64.4 & 0.0 & 0.0 & 0.1 & 0.0 & 0.1 & 24,941 & 10,953 & 4 \\
205 & 35 & 50 & 672.01 & 0.6 & 2.4 & 3.6 & 1.4 & 120 & 100.0 & 166.7 & 0.0 & 67.5 & 0.0 & 0.0 & 2.3 & 0.0 & 0.0 & 60 & 24 & 5 \\
206 & 35 & 60 & 626.66 & 0.3 & 4.2 & 4.4 & 4.4 & 252 & 98.8 & 34.1 & 0.0 & 0.0 & 0.0 & 0.0 & 4.3 & 0.0 & 0.0 & 1,384 & 293 & 4 \\
207 & 35 & 60 & 688.44 & 0.0 & 1.0 & 2.7 & 2.7 & 319 & 100.0 & 65.2 & 0.0 & 0.0 & 0.1 & 0.2 & 0.3 & 0.0 & 0.0 & 16 & 0 & 4 \\
208 & 35 & 60 & 615.54 & 1.2 & 3.6 & 7.3 & 7.0 & 418 & 49.8 & 99.0 & 21.3 & 17.5 & 0.0 & 0.0 & 1.9 & 2.9 & 0.0 & 11,933 & 477 & 3 \\
210 & 35 & 60 & 604.65 & 1.1 & 0.1 & 6.0 & 3.4 & 355 & 59.7 & 101.4 & 0.0 & 23.4 & 0.0 & 0.0 & 0.0 & 0.0 & 0.1 & 926 & 307 & 70 \\
211 & 35 & 70 & 582.31 & 2.3 & 2.9 & 8.6 & 8.3 & 473 & 71.0 & 74.0 & 0.0 & 19.2 & 0.0 & 0.0 & 1.4 & 0.0 & 0.0 & 12,182 & 914 & 2 \\
212 & 35 & 70 & 639.77 & 0.0 & 2.6 & 2.6 & 2.6 & 269 & 100.0 & 0.0 & 0.0 & 0.0 & 0.0 & 0.0 & 0.0 & 0.0 & 0.0 & 133 & 8 & 3 \\
213 & 35 & 70 & 583.71 & 1.2 & 4.3 & 6.4 & 6.4 & 466 & 78.1 & 56.2 & 0.0 & 0.0 & 0.0 & 0.0 & 0.2 & 0.0 & 0.0 & 75,434 & 3,830 & 25 \\
215 & 35 & 70 & 535.41 & 0.8 & 1.4 & 7.0 & 5.0 & 258 & 114.0 & 125.6 & 0.0 & 64.3 & 0.0 & 0.0 & 1.6 & 0.0 & 0.1 & 7,696 & 2,288 & 3 \\
216 & 35 & 80 & 542.79 & 3.4 & 8.7 & 12.4 & 11.9 & 424 & 93.9 & 45.5 & 0.0 & 18.2 & 0.0 & 0.0 & 1.1 & 0.0 & 0.0 & 79,852 & 7,956 & 4 \\
217 & 35 & 80 & 587.12 & 2.7 & 3.8 & 4.5 & 4.4 & 166 & 104.8 & 31.3 & 0.0 & 32.5 & 0.0 & 0.0 & 0.2 & 0.0 & 0.2 & 1,192 & 64 & 3 \\
220 & 35 & 80 & 478.64 & 1.8 & 4.4 & 11.5 & 9.4 & 246 & 88.6 & 89.0 & 0.0 & 28.9 & 0.0 & 0.0 & 18.2 & 0.0 & 0.1 & 35,438 & 13,561 & 2 \\ [0.3em]
221 & 40 & 50 & 780.25 & 0.3 & 6.3 & 7.1 & 3.8 & 392 & 100.5 & 35.5 & 0.0 & 32.1 & 0.0 & 0.0 & 8.3 & 0.0 & 0.0 & 1,224 & 116 & 3 \\
222 & 40 & 50 & 891.71 & 0.0 & 0.3 & 0.9 & 0.7 & 295 & 100.0 & 75.9 & 0.0 & 12.2 & 0.1 & 0.2 & 1.3 & 0.0 & 0.2 & 1 & 0 & 3 \\
223 & 40 & 50 & 788.17 & 0.2 & 1.9 & 4.6 & 4.5 & 240 & 80.0 & 79.2 & 0.0 & 26.7 & 0.0 & 0.0 & 1.1 & 0.0 & 0.0 & 5,263 & 1,063 & 2 \\
224 & 40 & 50 & 1,203.44 & 5.2 & 6.2 & 9.8 & 7.6 & 286 & 129.7 & 260.1 & 0.0 & 65.7 & 0.0 & 0.0 & 0.0 & 0.0 & 0.0 & 64,483 & 26,793 & 1,745 \\
225 & 40 & 50 & 853.92 & 0.1 & 1.1 & 5.3 & 2.2 & 314 & 105.4 & 104.1 & 0.0 & 100.3 & 0.0 & 0.0 & 0.1 & 0.0 & 0.1 & 213 & 67 & 9 \\
226 & 40 & 60 & 729.51 & 0.7 & 4.4 & 6.9 & 6.7 & 561 & 85.6 & 52.4 & 0.0 & 18.2 & 0.0 & 0.0 & 3.2 & 0.0 & 0.0 & 12,993 & 1,119 & 3 \\
227 & 40 & 60 & 767.96 & 0.0 & 2.3 & 4.1 & 3.1 & 318 & 100.0 & 42.5 & 0.0 & 42.8 & 0.0 & 0.1 & 0.1 & 0.0 & 0.6 & 46 & 4 & 2 \\
228 & 40 & 60 & 727.64 & 2.2 & 7.8 & 9.7 & 9.7 & 673 & 81.4 & 54.5 & 0.0 & 0.0 & 0.0 & 0.0 & 5.8 & 0.0 & 0.0 & 62,686 & 4,244 & 3 \\
230 & 40 & 60 & 726.17 & 1.4 & 4.0 & 8.6 & 4.9 & 294 & 91.2 & 122.8 & 0.0 & 39.8 & 0.0 & 0.0 & 2.2 & 0.0 & 0.1 & 3,225 & 620 & 3 \\
231 & 40 & 70 & 644.21 & 3.4 & 9.1 & 11.9 & 10.6 & 412 & 94.9 & 18.2 & 0.0 & 42.5 & 0.0 & 0.0 & 11.1 & 0.0 & 0.0 & 81,130 & 7,755 & 3 \\
232 & 40 & 70 & 718.35 & 1.0 & 3.7 & 4.8 & 4.8 & 264 & 117.0 & 56.1 & 0.0 & 0.0 & 0.0 & 0.0 & 0.3 & 0.0 & 0.0 & 569 & 40 & 1 \\
235 & 40 & 70 & 647.48 & 0.7 & 3.0 & 4.4 & 4.4 & 326 & 106.4 & 89.6 & 0.0 & 0.0 & 0.0 & 0.0 & 2.2 & 0.0 & 0.0 & 19,735 & 6,834 & 3 \\
237 & 40 & 80 & 660.41 & 0.0 & 4.1 & 4.1 & 4.1 & 432 & 73.6 & 0.0 & 0.0 & 0.0 & 0.0 & 0.0 & 0.0 & 0.0 & 0.0 & 4,195 & 347 & 3 \\ \midrule
\multicolumn{3}{r}{\bf min} &   & 0.0  & 0.0 & 0.0 & 0.0 & & 0.0 & 0.0 & 0.0 & 0.0  & 0.0 & 0.0 & 0.0 & 0.0 & 0.0  & 0  & 0 & 0 \\
\multicolumn{3}{r}{\bf max}   &   & 11.1 & 23.1 & 28.6 & 22.4  &  &  278.2 & 291.4 & 180.0 & 112.8 & 0.5 & 0.8 & 100.0 & 13.3 & 2.4  & 138,571  & 32,635 & 2,050 \\
\multicolumn{3}{r}{\bf mean}    &  & 1.4 & 3.5 & 7.8 & 6.7  &  & 90.8 & 80.9 & 5.6 & 28.0 & 0.0 & 0.1 & 5.7 & 0.6 & 0.2 & 9,185 & 1,200 & 45 \\
\multicolumn{3}{r}{\bf 25\%-quantile}   &   & 0.0 & 1.0 & 2.9 & 2.2 &  & 79.6 & 44.9 & 0.0 & 0.0 & 0.0 & 0.0  & 0.1 & 0.0 & 0.0 & 3  & 0 & 1  \\
\multicolumn{3}{r}{\bf median}     &   & 0.8 & 2.8 & 6.3 & 5.0 &  & 99.8 & 87.4 & 0.0 & 18.8 & 0.0 & 0.0  & 1.2 & 0.0 & 0.0 & 152  & 9 & 3  \\
\multicolumn{3}{r}{\bf 75\%-quantile}     &   & 2.4 & 4.9 & 11.9 & 9.7 &  & 100.0 & 106.7 & 0.0 & 46.2 & 0.0 & 0.1  & 4.5 & 0.2 & 0.0 & 4,203  & 313 & 4  \\  \bottomrule
\end{longtable}

\end{landscape}
\end{document}